\documentclass[sigplan,screen]{acmart} 

\usepackage[capitalise]{cleveref}
\usepackage{multirow}
\usepackage{amsmath}
\usepackage{mathtools}
\usepackage{algorithm}
\usepackage{algorithmicx}
\usepackage{algpseudocode}
\usepackage{setspace}

\usepackage{tikz}
\usepackage{balance}

\AtBeginDocument{%
  }

\renewcommand\footnotetextcopyrightpermission[1]{}
\acmDOI{10.1145/3696443.3708937}
\acmYear{2025}
\copyrightyear{2025}
\acmISBN{979-8-4007-1275-3/25/03}
\acmConference[CGO '25]{Proceedings of the 23rd ACM/IEEE International Symposium on Code Generation and Optimization}{March 01--05, 2025}{Las Vegas, NV, USA}
\acmBooktitle{Proceedings of the 23rd ACM/IEEE International Symposium on Code Generation and Optimization (CGO '25), March 01--05, 2025, Las Vegas, NV, USA}
\begin{document}

\title{Mantra: Rewriting Quantum Programs to Minimize Trap-Movements for Zoned Rydberg Atom Arrays}

\author{Enhyeok Jang}
\orcid{0009-0000-7034-6793}
\affiliation{%
  \institution{Yonsei University}
  \city{Seoul}
  \country{Republic of Korea}
}
\email{enhyeok.jang@yonsei.ac.kr}

\author{Youngmin Kim}
\orcid{0009-0002-8346-4830}
\affiliation{%
  \institution{Yonsei University}
  \city{Seoul}
  \country{Republic of Korea}
}
\email{youngmin.kim@yonsei.ac.kr}

\author{Hyungseok Kim}
\orcid{0009-0003-8981-3015}
\affiliation{%
  \institution{Yonsei University}
  \city{Seoul}
  \country{Republic of Korea}
}
\email{kimnumber@yonsei.ac.kr}

\author{Seungwoo Choi}
\orcid{0009-0005-2162-8993}
\affiliation{%
  \institution{Yonsei University}
  \city{Seoul}
  \country{Republic of Korea}
}
\email{seungwoo.choi@yonsei.ac.kr}

\author{Yipeng Huang}
\orcid{0000-0003-3171-6901}
\affiliation{%
  \institution{Rutgers University}
  \city{New Brunswick}
  \country{USA}
}
\email{yipeng.huang@rutgers.edu}

\author{Won Woo Ro}
\orcid{0000-0001-5390-6445}
\affiliation{%
  \institution{Yonsei University}
  \city{Seoul}
  \country{Republic of Korea}
}
\email{wro@yonsei.ac.kr}



\begin{abstract}

A zoned neutral atom architecture achieves exceptional fidelity by segregating the execution spaces of 1- and 2-qubit gates, being a promising candidate for high-accuracy quantum systems. 
Unfortunately, naively applying programs designed for static qubit topologies to zoned architectures may result in most execution time being consumed by inter-zone travels of atoms.
To address this, we introduce \textit{Mantra} (Minimizing trAp movemeNts for aTom aRray Architectures), which rewrites quantum programs to reduce the interleaving of single- and two-qubit gates. 
\textit{Mantra} incorporates three strategies: (i) a fountain-shaped controlled-Z (CZ) chain, (ii) ZZ-interaction protocol without a 1-qubit gate, and (iii) preemptive gate scheduling.
\textit{Mantra} reduces inter-zone movements by 68\%, physical gate counts by 35\%, and improves circuit fidelities by 17\% compared to the standard executions.


\end{abstract}








\maketitle

\section{Introduction}

\begin{figure} [t] 
  \centerline {
  \includegraphics [width=\columnwidth] {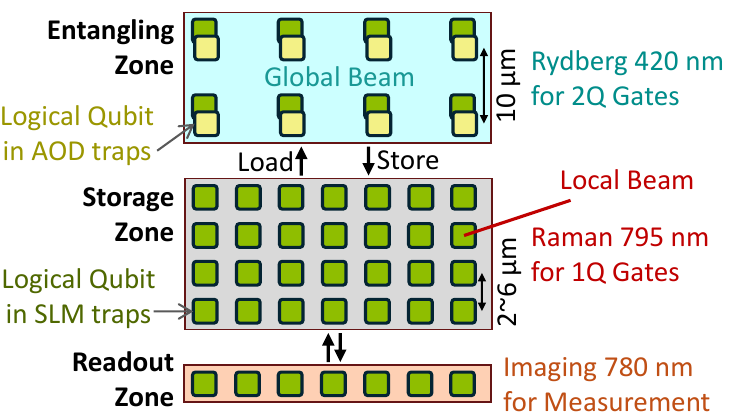} }
  \caption {
    The zoned architecture consisting of logical qubits is divided into three zones: entangling (for 2-qubit gate executions), storage (for 1-qubit gate executions), and readout (for measurement) zones \cite{bluvstein2024logical}.
    Some qubits are stored in the static optical tweezer array generated by the spatial light modulator (SLM).
    Some other qubits are in the movable tweezer array generated by the two-dimensional acousto-optic deflectors (AODs).
    \texttt{Load} means that the logical qubit trapped by the AOD moves from the storage to the entangling zone, and \texttt{Store} means from the entangling to the storage zone.
  } 
  \Description[<short description>]{<long description>}
  \label{f1} 
\end{figure}

Quantum computers utilizing neutral atoms manipulated by optical tweezer arrays can offer a scalable architecture thanks to their flexible reconfigurability, long coherence time of qubits, and parallel gate execution capabilities \cite{bauer2024solving, graham2022multi, finvzgar2024quantum, scholl2023erasure}. 
One of the recent advancements in this domain is the zoned neutral atom-based qubit architecture, which significantly enhances the fidelity of quantum gates by separating the execution spaces between 1-qubit gates and 2-qubit gates, as shown in \cref{f1} \cite{sunami2024scalable, stade2024abstract, quemix, claes2022high, li2024high, ball2024, stade2024optimalstatepreparationlogical}. 
The zoned architecture has reported 1-qubit and 2-qubit gate fidelities exceeding 99.9\% and 99.5\%, respectively, surpassing the performance of previously known neutral atom qubit architectures \cite{bluvstein2024logical, evered2023high}. 

Despite the superior fidelity, the zoned architecture may introduce an unprecedented challenge related to zone-to-zone movements of atoms. 
While the pulse application time for gate execution is less than a single microsecond \cite{evered2023high, wang2024atomique}, the atom's traveling between zones can require hundreds of microseconds  \cite{bluvstein2024logical, stade2024abstract}. 
Note that the trap travel speed is limited to about 0.55 $\mu$m/$\mu$s in order not to miss atoms trapped \cite{evered2023high}.
This difference in time scales implies that movements due to the requirement to relocate qubits to another zone for different kinds of gate execution may be a major execution bottleneck of quantum programs.
Our experiments with various quantum program benchmark \cite{quetschlich2023mqt, li2023qasmbench, tomesh2022supermarq} reveal that a na\"ive program execution on zoned architectures may result in an average of 78.2\% (and up to 89.9\%) of the total execution time being consumed by zone-to-zone movements of qubits.

We note that gate arrangements in quantum program structures commonly observed may not be optimal in zoned architectures.
Applying them to zoned architectures may introduce frequent switching between single-qubit and two-qubit gate operations.
For example, translating the UCCSD (Unitary coupled-cluster single and double \cite{li2022paulihedral}) circuits for the molecular chemical simulations \cite{jin2024tetris} into representations that zoned architectures can execute requires frequently switching of the execution of 2-qubit CZ gates and single-qubit Hadamard gates.
Programs such as quantum neural networks (QNNs \cite{markidis2023programming, mcclean2018barren, bermejo2024quantumconvolutionalneuralnetworks}) or quantum approximation optimization algorithms (QAOA \cite{farhi2014quantum, jang2024recompiling}), where locally entangled gate structures based on ZZ-interactions are prevalent, also require frequent switching between 2-qubit CZ and single-qubit RX or Hadamard gates \cite{nielsen2010quantum} in zoned architecture execution.
These interleavings of gates could increase inter-zone movements of atoms, resulting in longer runtimes.

Unfortunately, most previous quantum compiler studies on neutral atom quantum processors are developed for non-zoned architectures \cite{li2023timing, patel2022geyser, tan2024compiling, wang2024q, wang2024atomique, baker2021exploiting, brandhofer2021optimal, schmid2023hybrid, tan2024depth} or mainly minimizes intra-zone movements for the zoned architecture \cite{stade2024abstract}.
Thus, only leveraging existing compilation techniques may not sufficiently reduce the qubits' zone-to-zone movement overhead.
By rewriting \cite{kommrusch2023self, floyd1964bounded, johnson2022martini} quantum programs to mitigate frequent transitions between single-qubit and two-qubit gate execution, the inter-zone movement overhead in zoned architectures would be reduced. 
We introduce \textit{Mantra} (Minimizing trAp movemeNts for aTom aRray Architectures), a quantum program generation technique designed to reduce zone-to-zone movements in zoned architectures.

\textit{Mantra} reduces the number of necessary single-qubit gate operations in the storage zone; and, when the movement of qubits between zones is unavoidable, \textit{Mantra} greedily runs gates in advance of inter-zone travel.
First, \textit{Mantra} adopts a fountain-shaped CZ-tree structure, which allows the cancellation of the single-qubit gates between CZs when translating the molecular simulation circuits into a CZ gate-based circuit.
This CZ-tree structure eliminates the requirement for zoned quantum processors to send some logical qubits to the storage zone while performing CZ-tree operations.
This tree structure is optimal for running the transversal gate from a trap transfer overhead perspective, as only one logical qubit is moved by AOD, and the other qubits remain in the SLM trap.
In addition, \textit{Mantra} introduces a ZZ-interaction control protocol consisting of only Rydberg-mediated gates without a single-qubit gate.
This protocol eliminates the need for some logical qubits to go to the storage zone while processing serial ZZ rotational operations, such as the cost Hamiltonian in the QAOA circuits.
Finally, \textit{Mantra} reduces the number of inter-zone movements through the preemptive execution of other gates with no computational dependencies but should be run in the same zone. 
Fetching and aligning these gates in advance, \textit{Mantra} reduces inter-zone travels of logical qubits.

According to the evaluations using various benchmarks \cite{quetschlich2023mqt, li2023qasmbench, tomesh2022supermarq}, \textit{Mantra} achieves an 88\% reduction in inter-zone movements, a 35\% decrease of physical gate, and a 17\% improvement of fidelity compared to standard executions. 
Applying \textit{Mantra} to the zoned architecture yields an average execution time that is 57\% and 41\% shorter than that of state-of-the-art compilers for the general-purpose neutral atom architectures \cite{wang2024atomique} and zoned architectures \cite{stade2024abstract}, respectively.

We also confirm the benefits of the program rewriting strategies by \textit{Mantra} in scenarios where applying Raman laser for 1-qubit gate operation is permitted directly within the entangling zone. 
In such cases, the atom transfer overhead between AOD and SLM traps emerges as a new execution bottleneck, as atoms do not need to move to the storage zone for each gate interleaving. 
Our evaluation across 7 quantum benchmarks shows that \textit{Mantra} effectively mitigates trap transfer and AOD shuttling overhead, reducing overall execution times by up to 87\% and an average of 57\%.

By considering the challenges of designing compilers for zoned neutral atom processors, \textit{Mantra} can provide guidance for higher-level quantum developers writing programs for such architectures. 
In doing so, \textit{Mantra} can bridge the gap in the computing stack between high-level quantum algorithm design and the physical program execution in zoned systems. 
For instance, while algorithm designers often default to using CX gates, zoned architectures can leverage a variety of Rydberg-mediated gates, such as CZ, CPhase, or LP gates, which offer additional flexibility. 
Moreover, high-level developers might overlook the costs of trap transfers and zone-to-zone movements caused by alternating between single- and two-qubit gates.
\textit{Mantra} can mitigate these inefficiencies by exploiting the flexibility of CZ-tree chain synthesis, various Rydberg-mediated gates, and proactive scheduling of gates.

The applying coverage of methodologies by \textit{Mantra} includes both near-term quantum applications \cite{preskill2018quantum} without quantum error correction (QEC \cite{steane1996multiple, calderbank1996good}) and fault-tolerant quantum applications considering QEC.
Our proposed program rewriting methodologies could be applied to physical qubits without QEC, and they could also be easily extended to the identical principle for logical qubits consisting of several physical qubits that perform the same operations in parallel, such as Shor or Steane state correction codes \cite{calderbank1996good}.

The five main contributions of this paper are as follows.
\begin{itemize}
    \item We observe that the inter-zone movements of qubits in zoned neutral atom architectures could be a major run-time bottleneck for quantum program executions.
    \item We also note that frequently interleaved executions between the single-qubit gate and two-qubit gate observed in various quantum program structures could increase the zone-to-zone movements of logical qubits.
    \item The proposed method, \textit{Mantra}, exploits three main techniques, including an efficient CZ chain structure, a new arbitrary ZZ rotation protocol, and a preemptive gate alignment to reduce the inter-zone movements.
    \item \textit{Mantra} reduces the number of cross-zone movements by 68\% compared to standard program execution. 
    In addition, \textit{Mantra} can generate quantum programs that provide 57\% and 41\% less run-time compared to most recent compilers for general-purpose atom array-based architectures and the zoned architecture, respectively.
    \item \textit{Mantra} mitigates the AOD-SLM trap transfer and shuttling overhead, thereby reducing execution times by up to 87\% and an average of 57\% even in scenarios where Raman laser is allowed in the entangling zone.
\end{itemize}

\section{Background and Motivation}

This section introduces a zoned neutral atom architecture where 1-qubit and 2-qubit gate execution spaces are isolated. 
In this architecture, programming is achieved through pulse-level controls and the movement of atoms using optical traps. 

Next, we examine the implementation of SWAP operations in zoned architectures. 
In atom-based quantum computers, SWAP can be realized either through gate-based operations or by directly repositioning atoms. 
We highlight the inefficiency of gate-based SWAPs in zoned architectures due to substantial inter-zone movement overheads, underscoring the importance of prioritizing qubit moving-based SWAPs.

Finally, we analyze the execution time of various quantum benchmarks, demonstrating that inter-zone movements may dominate run-time in zoned architectures. 
This observation motivates the development of new compilation techniques specifically designed to optimize zone-to-zone movements, as opposed to those tailored for the non-zoned architecture.

\subsection{Zoned Neutral Atom Array Architecture}

Zoned architecture is designed to enhance fidelity by segregating spaces between 1-qubit and 2-qubit gate executions.

\textbf{Hardware Configuration}:
\cref{f1} illustrates the zoned architecture, consisting of 3 zones: entanglement, storage, and readout.
The entangling zone is dedicated to executing 2-qubit gates (e.g., CZ), while the storage zone is for single-qubit gate (e.g., single-qubit rotation gates) executions.
The readout zone is used for measuring qubits.
We consider a scenario for preparing fault-tolerant qubits using the Steane code, which is detailed in \cref{appendixa}.
Qubits are stored in the static optical tweezer array generated by the spatial light modulator (SLM) \cite{choi2023preparing}.
The movable tweezer array generated by the 2-dimensional accousto-optic deflectors (AODs) can carry qubits between zones \cite{evered2023high}.
For the convenience of explanation, we leverages the terms of \texttt{Load} and \texttt{Store} to refer to the movement of atoms by the AOD trap between the entanglement and storage zones \cite{stade2024abstract}.
\texttt{Load} instruction moves the logical qubit trapped by the AOD trap from storage to entangling zone, while \texttt{Store} instruction does the opposite.

\textbf{Programming by Pulse Controls and Trap Movements}:
The implementation of 2-qubit gates (e.g., transversal CZ) is realized by the Van der Waals' (dipole-dipole) interaction force between two neutral atoms close to each other (about 2 $\mu$m) \cite{bluvstein2022quantum}. 
This implementation is performing by a global pulse in the entangling zone that excites atoms to Rydberg states \cite{bluvstein2024logical}.
The implementation of single-qubit gates (e.g., single-qubit rotations) is implemented using Raman excitation through a AOD trap in the storage zone \cite{bluvstein2024logical}.
The readout process is enabled by moving qubits to the readout zone and illuminating with the focused imaging beam \cite{evered2023high}.

\subsection{SWAP Implementation on Zoned Architectures} \label{swap}

\begin{figure} [h] 
  \centerline {
  \includegraphics [width=\columnwidth] {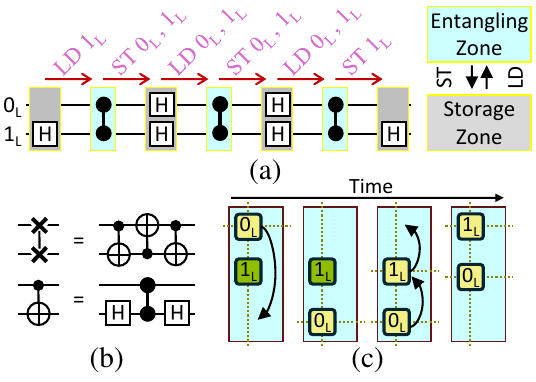} }
  \caption {
    (a) An implementation of gate-based SWAP on the zoned architecture, where LD stands for \texttt{Load} and ST stands for \texttt{Store}. 
    $L$s within $0_L$ and $1_L$ mean that they are logical qubits encoded into seven physical qubits by the Steane code \cite{steane1996multiple}, respectively.
    (b) The equivalent gate decompositions for explaining CZ gate-based SWAP operation in (a): a single SWAP gate could be rewritten with 3 CXs
    and a single CX could be rewritten with two Hadamard gates and a CZ \cite{nielsen2010quantum}. 
    (c) An implementation of the movement-based SWAP that complies with the movement constraints of atoms \cite{tan2024compiling, bluvstein2024logical}.
  } 
  \Description[<short description>]{<long description>}
  \label{f2} 
\end{figure}

SWAP operation is essential for quantum programming to exchange information between 2 qubits with each other \cite{nielsen2010quantum}.
On fixed qubit architectures, SWAP is required to perform a two-qubit gate operation between two specific qubits unconnected topologically \cite{ibm}.
Since neutral atom-based qubits can be movable directly, SWAP can be implemented in two ways: (i) by applying pulses that realize SWAP operation, as is done in fixed qubit architectures (gate-based), and (ii) by exchanging positions of qubits within traps (trap-movement based) \cite{schmid2024computational}.
Two SWAP implementation methods have distinct advantages, and the latest quantum compilers for non-zoned architectures often employ a hybrid approach, selecting the appropriate SWAP type based on the context \cite{wang2024atomique, tan2024compiling}. 
In contrast, we observe that trap movement-based SWAPs might be superior to gate-based SWAPs for zoned architectures in all scenarios.
Gate-based SWAPs in zoned architectures require six inter-zone movements (illustrated in \cref{f2} (a)), leading to unnecessary execution overhead. 
The detailed gate decomposition process is shown in \cref{f2} (b).  

When a two-qubit entangling operation should be processed by moving qubits from the storage zone to the entangling zone, the requirements of SWAP are addressed naturally, as it is a remapping of logical qubits on the physical qubits in the storage zone.
Even when a SWAP is required for qubits within the entangling zone, sequential movement adhering to the constraints of AOD trap movement \cite{stade2024abstract}, as shown in \cref{f2} (c), is more efficient than a gate-based SWAP.

Note that implementing a gate-based SWAP exclusively in the entangling zone is not allowed due to the requirement of 1-qubit Hadamard gates.  
This implies that the gate-based SWAP may offer no advantage over trap movement-based ones as long as single- and two-qubit gate executions remain segregated. 
Therefore, we would adopt the trap movement-based SWAP method with priority for zoned architectures.



\subsection{Breakdown Analysis on Zoned Architecture}

\begin{figure} [h] 
  \centerline {
  \includegraphics [width=\columnwidth] {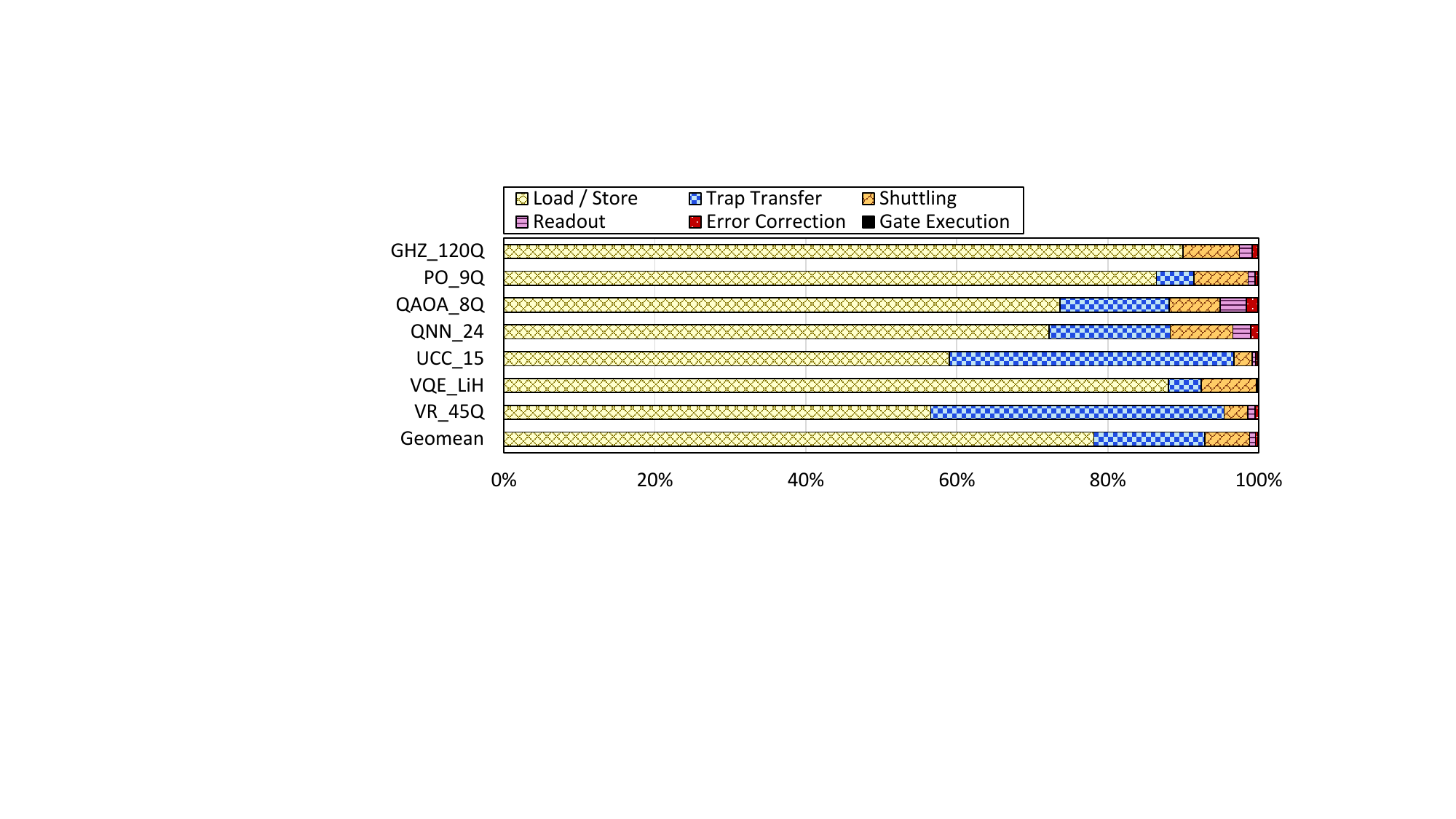} }
  \caption {
    Execution time breakdown on zoned architectures. 
    \texttt{Load/Store} is the time for qubits to move to another zone.
    \texttt{Trap Transfer} is the time to handover a qubits between a movable AOD and a static SLM trap.
    \texttt{Shuttling} is the time qubits spend moving in the entangling zone.
    \texttt{Readout} is the time required for measuring.
    \texttt{Error Correction} is the time spent preparing logical qubits and detecting errors.
    \texttt{Gate Execution} is the pulse application time for executing gates.
  } 
  \Description[<short description>]{<long description>}
  \label{f3} 
\end{figure}

\noindent
We perform the execution time breakdown analysis on the zoned architecture using various scalable quantum program benchmarks \cite{quetschlich2023mqt, tomesh2022supermarq, li2023qasmbench} as shown in \cref{f3}.
The major execution bottleneck of quantum programs on the zoned architecture is the inter-zone movements by Loads and Stores, which consumes an average of 78.2\% and up to 89.9\% of the total execution time.
Zoned architectures pose a new bottleneck of the inter-zone movements, unlike non-zoned architectures.

Note that the trap transfer and shuttling is often the main execution time overhead in non-zoned neutral atom architectures \cite{tan2024depth, tan2024compiling, decker2024arctic}.
Therefore, various compiler techniques have been proposed for the initial mapping and routing of logical qubits to mitigate the trap transfer and shuttling overhead \cite{li2023timing, patel2022geyser, wang2024q, wang2024atomique, baker2021exploiting, brandhofer2021optimal, schmid2023hybrid}.
However, according to our evaluation, time for the trap transfer and shuttling account for less than 20\% of the total runtime in zoned architectures.
This suggests that previous compilation techniques developed for non-zoned architectures may not contribute significantly to reducing execution time for zoned architectures, which will be covered in more detail in the evaluation section.
In other words, it is motivated by the need to develop compilation techniques different from the existing ones to optimize the zone-to-zone movements of qubits for zoned architectures.

\section{Mantra Compilation Methodology}

This section describes the compile methodology of \textit{Mantra} to reduce the inter-zone movement in the zoned architectures.
We note that the frequent switching of single-qubit and two-qubit gate executions can occur when quantum programs run na\"ively on zoned neutral atom architectures.
One may wonder why quantum program structures are written to require these frequent interleavings of gate execution.
This may result from the high-level quantum programs (or even from the quantum algorithm) being written by adopting the CX gate as a representative two-qubit gate.
Note that CX gates in quantum programs to execute on neutral atom-based processors should be translated into the representation based on the CZ and H (i.e., their native gates).
Due to this translation process, 1-qubit and 2-qubit gate executions could be frequently interleaved, resulting in excessive zone-to-zone movements of atoms.
The execution efficiency of the zoned architectures could be improved by rewriting programs so that they require fewer zone-to-zone movements of qubits.

\textit{Mantra} minimizes 1-qubit gate operations required for the storage zone execution.
\textit{Mantra} adjusts program scheduling to preemptively run some gates that could be executed in the same zone when the inter-zone movement is inevitable.

\subsection{Fountain-Shaped CZ-Tree Chain} \label{fountain}

\begin{figure} [h] 
  \centerline {
  \includegraphics [width=\columnwidth] {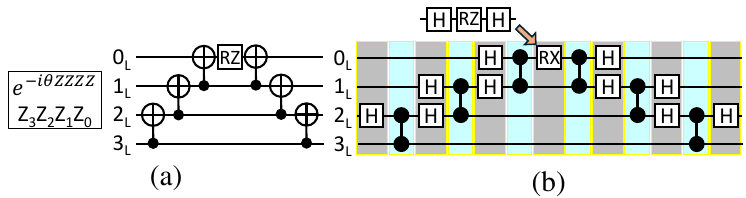} }
  \caption {
    (a) An example Hamiltonian simulation kernel \cite{li2022paulihedral} and its circuit implementation with standard chain structure, and
    (b) CZ Gate-Based Implementation of the circuit of (a).
  } 
  \Description[<short description>]{<long description>}
  \label{f4} 
\end{figure}

\noindent
This section discusses the efficiently executed CZ-tree chain structure on the zoned neutral atom architectures, where the chain structure is required for Hamiltonian (or quantum) simulation kernels \cite{li2022paulihedral} such as UCCSD (Unitary Coupled-Cluster Single and Double \cite{barkoutsos2018quantum, sokolov2020quantum}) circuits.
\cref{f4} (a) shows an example Pauli string and its corresponding quantum circuit, where the CX gates are generally constructed in a path-shaped standard chain structure, where the control qubit of the next CX becomes the target qubit of the previous CX.
Unfortunately, as shown in \cref{f4} (b), the CZ gate-based circuit translated from \cref{f4} (a) may raise frequent switchings between the execution of single-qubit and 2-qubit gates.
Since the circuit in \cref{f4} (b) requires single-qubit Hadamard (H \cite{nielsen2010quantum}) or rotation-X (RX \cite{nielsen2010quantum}) gate(s) between each CZ gate execution, inter-zone movement is required for each gate execution step.
For example, a $n$-qubit single Pauli-string circuit could require at least $2(n-1)$ times of interleavings for the different zone execution in the standard chain structure.

\begin{figure} [h] 
  \centerline {
  \includegraphics [width=\columnwidth] {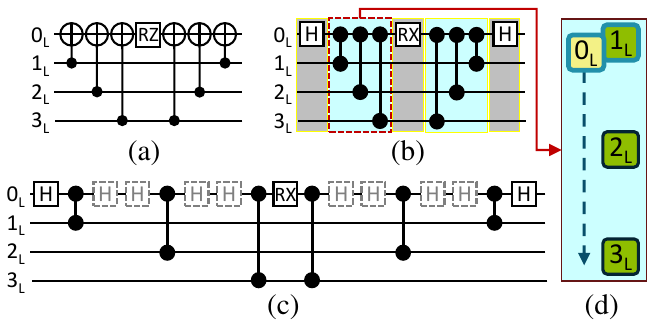} }
  \caption {
    (a) Implementation of a circuit with fountain-shaped chain structure for the simulation kernel in \cref{f4},
    (b) CZ-based Implementation of the circuit of (a),
    (c) Describing the gate cancellation process of H gates in the fountain CZ chain, and
    (d) A possible AOD trap movement for CZ chain.
  } 
  \Description[<short description>]{<long description>}
  \label{f5} 
\end{figure}

The (path-shaped) standard chain structure of executed in the zoned architecture may result in frequent inter-zone movement.
To address this, \textit{Mantra} adopts a fountain-shaped chain structure, as shown in \cref{f5} (a).
As shown in \cref{f5} (b), the target qubit of all CX gates in the proposed fountain-shaped chain structure is $0_L$ in common.
Note that two consecutive Hadamard gates become the identity operation since the Hadamard operator is a Hermitian matrix \cite{noble1977applied}.
As described in \cref{f5} (c), the CZ trees are only processed within the entangling zone without having to travel to the storage zone since Hadamard gates between CZs are canceled.
Therefore, the fountain-shaped chain structure could reduce inter-zone movements required for the CZ-tree executions.

This fountain-shaped CZ chain structure not only reduces the number of zone-to-zone movements for logical qubits but is also efficient from the perspective of atom shuttling within the entangling zone.
$0_L$ qubit participates in a computation for all CZ in the tree operations.
Therefore, transversal CZ operations could be achieved by only moving the logical qubit of $0_L$ by the AOD trap, as described in \cref{f5} (d).
The standard chain structure described in \cref{f4} requires frequent trap transfers between AOD and SLMs since they should be moved to the storage zone immediately after CZ operation. 
In contrast, the fountain-shaped chain structure does not require any AOD-SLMs trap transfer during a CZ-tree operation process.
Therefore, the fountain-shaped CZ chain structures are expected to be relatively more efficient than standard chain structures, not only in zoned but also in non-zoned architectures.
The cancellation of Hadamard gates in the CZ-tree reduces laser pulse application time slightly and thereby reduces fidelity degradation due to the single-qubit gate executions, and the trap transfers could also be reduced.


\begin{algorithm} [h] \footnotesize
\caption{Fountain-Shaped CZ-Tree Implementation}
\setstretch{0.98}
\begin{algorithmic}
\Function{cz\_fountain}{pauli: Pauli} 
    \State chain $\gets$ QC(pauli.num\_qubits)
    \State control, target $\gets$ None, None
    \State chain.h(target) //Applying Hadmamrd Gates
    \For{$i \gets 0$ to pauli.num\_qubits}
        \State pauli\_i $\gets$ pauli.to\_label[i]
        \If {pauli\_i $\neq$ `I'} 
            \If {target = \textbf{None}} 
                \State target $\gets$ i
            \Else 
                \State control $\gets$ i
            \EndIf
        \EndIf
        \If {control $\neq$ None \textbf{and} target $\neq$ None}
            \State chain.cz(control, target) //Applying Controlled-Z Gates
            \State control $\gets$ None
        \EndIf
    \EndFor
    \State chain.h(target)
    \State \textbf{return} chain
\EndFunction
\end{algorithmic} \label{a0}
\end{algorithm}

The fountain-shaped CZ chain structure shown in \cref{f5} (b) can be implemented as described in \cref{a0}.
Note that \textit{Mantra} is not the first to propose a fountain-shaped chain structure.
The flexibility of CX (or CZ) chain synthesis in Hamiltonian simulation kernels is well known \cite{li2022paulihedral, jin2024tetris}, and many commercial quantum compilers (such as IBM's Qiskit SDK \cite{ibm}) can provide various chain structures.
There are many recent studies for efficient placement of 2-qubit gate trees in sparsity topologies such as superconducting-based qubits \cite{jiang2020optimal, liu2024fermihedral, liu2024ternary, liu2024quclear}.
Furthermore, a method for synthesizing efficient fermionic system simulation circuits on non-zoned neutral atomic array processors has recently been proposed \cite{gonzalez2023fermionic}.
Unfortunately, we note that these recent studies cannot adequately reduce the inter-zone movement overhead, as they are not developed for zoned architectures.
In the implementation presented in \cref{a0}, the order of application of CZ is in ascending order of the qubit numbering due to the movement efficiency of AOD traps, as opposed to the descending-order provided by \texttt{Lie-Trotter} in Qiskit SDK.

\begin{figure} [h] 
  \centerline {
  \includegraphics [width=\columnwidth] {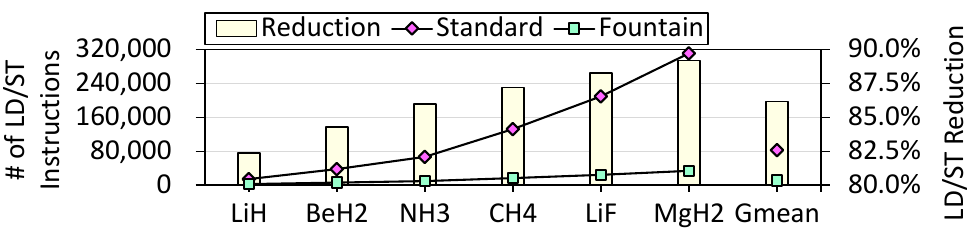} }
  \caption {
    Comparative analysis of the number of Load/Store instructions by two chain structures (Standard VS. Fountain) in VQE (Variational quantum eigensolver) circuits of the UCCSD (Unitary coupled-cluster single and double) Ansatz.
    In this experiment, all molecular simulation circuits were subjected to the gate cancellation by Qiskit Transpiler \cite{ibm}.
  } 
  \Description[<short description>]{<long description>}
  \label{f6} 
\end{figure}

We estimate the number of the LD and ST instructions according to the standard (path-shaped) and fountain-shaped chain structures for chemical simulation circuits encoded by the Jordan-Wigner transformation \cite{jordan1928paulische}, as described in \cref{f6}.
In the standard chain structure, the number of LD and ST instructions is proportional to four times the number of CZs in the simulation circuit, corresponding to complexity equivalent to $O(n^4)$ when the number of qubits is $n$.
On the other hand, the fountain-shaped chain structure requires two LDs and two STs for each Pauli-string (regardless of the number of qubits), resulting in lower complexity.
As a result, the LD and ST instruction counts in the fountain chain structure are more advantageously scaled than those in the standard chain.
As shown in \cref{f6}, the fountain chain structure requires 83\% fewer LD or ST instructions, averaging in six molecular simulation circuits over the standard chain.

\subsection{1Q-Gateless Arbitrary ZZ-Interaction Protocol}

\begin{figure} [h] 
  \centerline {
  \includegraphics [width=\columnwidth] {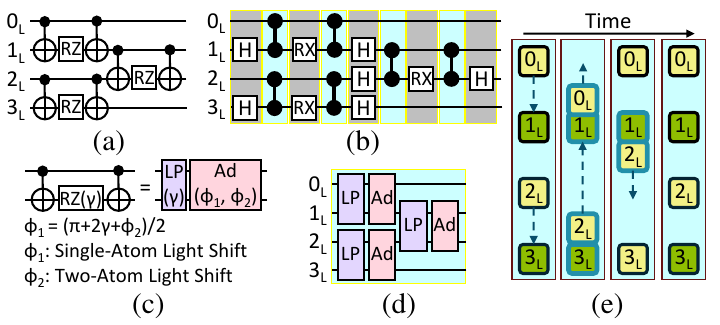} }
  \caption {
    (a) An example of the simulation circuit consisting of local entanglements with ZZ rotations,
    (b) CZ-based Implementation of the circuit of (a),
    (c) The proposed arbitrary ZZ rotation protocol consisting of a single adiabatic (Ad) \cite{keating2015robust} and a single Levine-Pichler (LP) gate \cite{levine2019parallel},
    (d) Implement the circuit in (a) by applying the proposed protocol, and
    (e) One possible qubit movement for circuit (d), where the entanglements could be implemented without both the AOD-SLM trap transfer and trip of logical qubits (to the storage zone).
  } 
  \Description[<short description>]{<long description>}
  \label{f7} 
\end{figure}

A ZZ-interaction is another prime operation for occurring interleaved executions of single- and two-qubit gates in the CZ gate-based decomposition template for zoned neutral atom architecture.
In this section, we analyze the execution efficiency of the zoned architecture in scenarios where locally entangled structures with arbitrary-angular ZZ interactions are prevalent across all logical qubit pairs, as shown in the \cref{f7} (a).
These locally entangled structures are often observed on the QAOA cost Hamiltonians for dense graphs \cite{farhi2014quantum} or ansätze for variational approach-based QNNs \cite{quetschlich2023mqt, amarasinghe2023variational}.

The circuit in \cref{f7} (a) is translated as \cref{f7} (b) based on the CZ gate to run on the zoned architecture.
Unfortunately, quantum program structures such as \cref{f7} (b) may cause excessive inter-zone movements of logical qubits due to frequent switching of 1-qubit and 2-qubit gate executions.
Moreover, in this program structure, it is challenging to apply the flexibility of the CZ synthesis \cite{li2022paulihedral} discussed in \cref{fountain}.
To address this challenge, \textit{Mantra} proposes a ZZ rotation gate protocol that does not require a single-qubit gate by leveraging two Rydberg-mediated gates capable of performing by neutral atoms.
The unitary operator for the arbitrary rotating ZZ gate to be implemented is as follows:
\begin{align*}
  RZZ(\gamma) &=
  \begin{bsmallmatrix}
        1 & 0 & 0 & 0 \\
        0 & e^{i\gamma} & 0 & 0 \\
        0 & 0 & e^{i\gamma} & 0 \\
        0 & 0 & 0 & 1 \\
  \end{bsmallmatrix}
\end{align*}
where $\gamma$ is a commonly used for cost Hamiltonians in QAOA.

The essential principle of the proposed gate protocol is to achieve the arbitrary angle ZZ-rotation by combining Levine-Pichler gates and CPhase (controlled phase) gates for offsetting phase shift for the $\lvert 11 \rangle$ state.
Hamiltonian for the optimal control of Rydberg atom-based qubits is described in detail in \cref{appendixb}.
There are various versions of entangling gates for realizing controlled phase shift \cite{evered2023high}, but in this section, we would describe them by adopting the adiabatic gate \cite{keating2015robust}.
Another version that implements the proposed gate protocol with the time-optimal \texttt{CPHASE} gate \cite{evered2023high} instead of the adiabatic gate is detailed in \cref{appendixc}.

The unitary operators implementing the adiabatic and the Levine-Pichler gate can be written as follows, respectively:
\begin{align*}
  Ad(\phi_{1}, \phi_{2}) =
  \begin{bsmallmatrix}
        1 & 0 & 0 & 0 \\
        0 & 1 & 0 & 0 \\
        0 & 0 & 1 & 0 \\
        0 & 0 & 0 & e^{\phi_{2}-2\phi_{1}} \\
  \end{bsmallmatrix}
  \text{ and }
  LP(\gamma) =
  \begin{bsmallmatrix}
        1 & 0 & 0 & 0 \\
        0 & e^{i\gamma} & 0 & 0 \\
        0 & 0 & e^{i\gamma} & 0 \\
        0 & 0 & 0 & e^{2\gamma+\pi} \\
 \end{bsmallmatrix}
\end{align*}
where $\phi_{1}$ could be selected by the single-atom light shift (which is induced by the global Rydberg laser \cite{mitra2020robust}), $\phi_{2}$ could be selected by the two-atom light shift \cite{keating2015robust}, and $\gamma$ is an angle for ZZ rotation.
By applying an adiabatic gate and a Levine-Pichler gate, where a single-atom and two-atom light shifts in the adiabatic gate are chosen so that $\phi_{1} = (\pi + 2\gamma + \phi_{2})/2$, the ZZ rotation could be achieved without any single-qubit gate, as shown in \cref{f7} (c).
It is attributed to the phase shift cancellation to the fourth amplitude ($\lvert 11 \rangle$) of the Levine-Pichler gate by controlled Z rotation from the adiabatic gate.

The proposed protocol does not require qubits to be sent to the storage zone while executing the ZZ-rotation operations, as it can be performed in the entangling zone only, as shown in \cref{f7} (d).
It differs from the program structure of \cref{f7} (b) that requires execution interleavings between different zones for every gate-layer.
One possible shuttling process without trap transfers can be realized as shown in \cref{f7} (e).


\subsection{Preemptive Identical-Zoned Gate Alignment} \label{preemptive}

\noindent
This section discusses a scheme to reduce the inter-zone movement overhead by adjusting the execution timing of some gates.
We utilize examples of GHZ-state preparation \cite{einstein1935can}.
The fundamental principle of the proposed gate alignment methodology can be described as follows:
First, find initial gate(s) in the program and identify its execution zone.
Next, among the subsequent gates in the program, the gates that have no computational dependency but can be processed together in the first zone execution step are identified.
These identified gates are preemptively aligned in the first zone execution step.
Then, identify gates that need to be processed in the next zone execution step and apply the identical rules.
This process is repeated until the final zone execution steps.

Similar to the case of Hamiltonian simulation circuits as discussed in \cref{fountain} \cite{cowtan2020generic, jin2024tetris}, there is a configuration flexibility of the CX (or CZ)-tree chain when preparing the GHZ state.
Connecting 2-qubit gates as the fountain shape may be more efficient rather than connecting CZs as the path shape.
The fountain-shaped CZ structure can reduce trap transfers and zone-to-zone travels as discussed in \cref{f5}.

\begin{figure} [h] 
  \centerline {
  \includegraphics [width=\columnwidth] {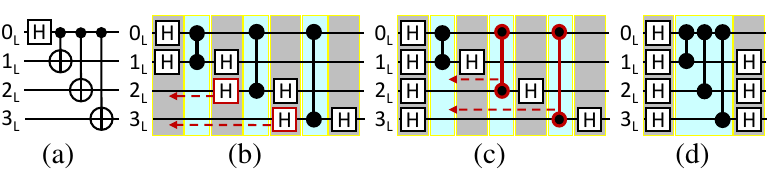} }
  \caption {
    (a) An example of a 4-qubit GHZ-state circuit consisting of the fountain-shaped CX chain,
    (b) CZ-based Implementation of the circuit of (a), where H gates first applied to the logical qubits $2_L$ and $3_L$ can be executed more preemptively before the first CZ.
    (c) The second and third CZs can be applied preemptively prior to the second storage zone execution, and
    (d) A circuit with the alignment applied.
  } 
  \Description[<short description>]{<long description>}
  \label{f8} 
\end{figure}

\cref{f8} (a) shows an example of a four-qubit GHZ state preparation circuit consisting of a fountain-shaped CX chain.
Note that their CXs' control and target qubits are opposite to the simulation circuits discussed in \cref{f5} when preparing GHZ states with fountain shape-based chains; the control qubit of CXs is concentrated in one qubit.
\cref{fountain} adopts a fountain-shaped CX-tree structure, exploiting the opportunity to cancel single-qubit Hadamard gates by sharing the target qubits of multiple CXs.
Unfortunately, in this scenario, the Hadamard gates would be generated from different qubits in the process of translating to CZ-based circuits, and hence, it is challenging to cancel them from each other.
Instead, in this scenario, we may consider an approach to adjust the execution schedule of some gates without computational dependence to reduce zone-to-zone movements of the qubit.

\cref{f8} (b) and (c) describe a preemptive gate alignment method to reduce the inter-zone travel overhead using an example of a 4-qubit fountain-shaped GHZ state circuit.
As shown in \cref{f8} (b), Hadamard gates initially applied to each of $2_L$ and $3_L$ can be preemptively executed in the first storage zone execution step.
Then, as shown in \cref{f8} (c), the CZ gates associated with $2_L$ and $3_L$ can also be preemptively executed in the first entangling zone execution step.
Applying this realignment provides circuits such as \cref{f8} (d), and it enables the preparation of GHZ status with only two inter-zone movements, regardless of the number of qubits.
If X-basis initialization and measurement are supported on the device, applying requirements for Hadamard gates are eliminated, so the GHZ state preparation could be performed without having to go to the storage zone during program execution.

\begin{figure} [h] 
  \centerline {
  \includegraphics [width=\columnwidth] {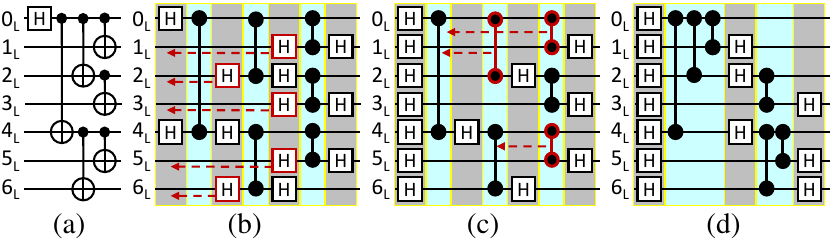} }
  \caption {
    (a) An example of a 7-qubit GHZ-state consisting of parallel CX chain \cite{cruz2019efficient},
    (b) CZ-based Implementation of the circuit of (a), where some H gates could be applied preemptively in the first step of the storage zone execution.
    (c) Some CZs could be applied in the entangling zone execution steps further ahead of their current location in the circuit, and
    (d) The circuit applied the preemptive gate alignment.
  } 
  \Description[<short description>]{<long description>}
  \label{f9} 
\end{figure}

Furthermore, we illustrate an example of reducing inter-zone travel overhead by using the proposed gate alignment method for the \textit{parallel} CX structure in which the GHZ-state circuit depth expands logarithmically as the number of qubits increases, as shown in \cref{f9} \cite{cruz2019efficient}.
\cref{f9} (a) shows an example of a 7-qubit GHZ state circuit consisting of the parallel CX chain.
Hadamard gates firstly applied to the logical qubits $1_L$, $2_L$, $3_L$, $5_L$, and $6_L$ could be preemptively executed in the first storage zone execution step, as shown in \cref{f9} (b).
Then, as shown in \cref{f9} (c), some CZs can also be preemptively executed in the earlier entangling zone execution step.
Applying this realignment provides circuits such as \cref{f9} (d).
In this case, the number of travels between the entangling zone and storage zone is reduced from 6 to 4.

\textbf{Comparison with Existing Compilation Techniques}:
Gate scheduling has been adopted for various purposes by many quantum program compilers \cite{qiskit2024, sivarajah2020t, BQSKit, stade2024abstract}, although to the best of our knowledge, there is no efficient one to reduce the inter-zone travel overhead in zoned architectures.
For example, \texttt{ASAPSchedule} and \texttt{ALAPSchedule} in Qiskit SDK \cite{qiskit2024} aims to reduce idling time by scheduling gates at the earliest or late possible.
Unfortunately, these schemes cannot be used to reduce atom movements between zones as they do not separate gates by 1-qubit and 2-qubit categories.

\subsection{Chain Execution Efficiency by Qubit Array Sizes} \label{efficiency}

This section analyzes the efficient CZ chain structure according to various sizes and topologies of logical qubit arrays in the zoned architecture, using GHZ state preparation as an example.
\cref{preemptive} described two CZ chain examples for GHZ state preparation in zoned architectures: fountain-shaped and parallel structures.
The parallel chains may seem more favorable for execution, as their circuit depth scales logarithmically with increasing qubits, while the fountain-shaped chain's one scales linearly.
However, regarding the number of inter-zone movements, the parallel chain requires logarithmically expanding LD or ST instructions as the number of qubits grows. 
In contrast, fountain-shaped chains require only a single LD and ST each regardless of the qubit counts.

Although the distance between the entangling and storage zone is only 20 $\mu$m \cite{bluvstein2024logical}, logical qubits (needed to go another zone) travel much further than that for the following reasons: 
(i) Each atom that makes up the qubit is separated from one another with sufficient distances on a two-dimensional lattice to avoid unintended interactions. 
Thus, as the qubit scale in a quantum program increases, the average travel distance of logical qubits for moving to different zones also grows.
(ii) Note that atoms in AOD traps cannot cross each other \cite{tan2024compiling, tan2024depth, stade2024abstract, wang2024atomique}.
This constraint on their movement can require some logical qubits to detour from other qubits or transfer others to SLM traps when they move to other zones.

\begin{figure} [h] 
  \centerline {
  \includegraphics [width=\columnwidth] {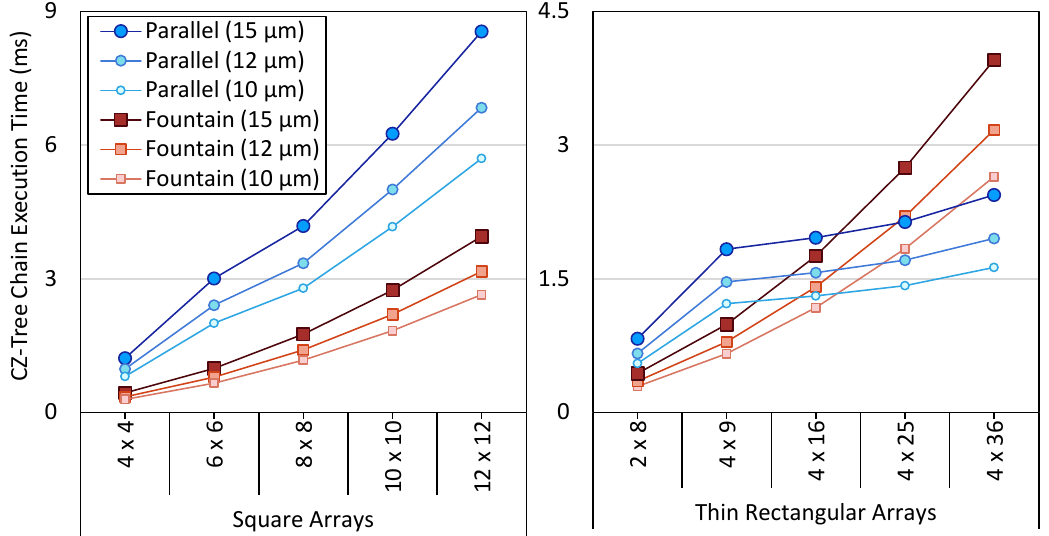} }
  \caption {
    A comparative analysis of the execution time of two CZ-tree chain structures (Parallel VS. Fountain) for preparing GHZ states by various 2-dimentional atom array configurations (square and thin rectangular-shaped) and inter-atom distances (10, 12, and 15 $\mu$m) in entangling zone.
  } 
  \Description[<short description>]{<long description>}
  \label{f10} 
\end{figure}

\cref{f10} compares the execution times of parallel-shaped and fountain-shaped chains for preparing GHZ states across various array sizes, shapes, and inter-atom distances. 
For square-shaped arrays, the fountain-shaped chain achieves shorter execution times on average compared to the parallel chain for the same number of qubits over the tested range of inter-atom distances and array sizes. 
Although parallel GHZ chains \cite{cruz2019efficient} are theoretically expected to be more efficient over fountain chains thanks to their logarithmic depth scalability when hardware topology is not considered, the results in \cref{f10} suggest this is not always true in practical hardware execution.
The difference in running time of the two chain structures is due to two reasons as follows:
(i) The application relationships and sequences of CZ gates in parallel chains are more complex than those in fountain-shaped chains, leading to linear scalability of atom travel time within the entangling zone as the number of qubits increases.
(ii) while the number of LD/ST operations in parallel chains scales logarithmically to the number of qubits, the fountain-shaped chain requires a constant number of LD/ST operations regardless of the number of qubits. 
As a result, the fountain-shaped chain shows better run-time scalability than the parallel chain in the 2-dimensional square arrays.

For thin rectangular-shaped arrays, which keep the number of the row of logical qubits constant, the average traveling time between zones would be kept almost constant.
In this case, the parallel CZ chain structure provides better time scalability with an increasing number of qubits compared to the fountain-shaped one, and at some point (approximately 80 physical qubits according to our calculation), the parallel chain would provide less execution time than the fountain-shaped chain.
Though the parallel CZ chain structure is more advantageous over the fountain-shaped chain in terms of execution time complexity, the practical CZ chain structure depends on qubit topologies (i.e., the arrangement of atoms).

\textbf{Discussion}:
\cref{f10} suggest that minimizing inter-zone travel distances of qubits through constraining the vertical size of zones (a thin rectangular array) could enhance execution efficiency, particularly for programs with high gate parallelism. 
Nevertheless, the thin rectangular array may limit the reconfigurability within zones. 
Instead, by expanding the contacted area between the entangling zone and storage zone, hardware developers can consider a design that, on average, allows large amounts of atom movement with short distances while maintaining reconfigurability within the zone.
For instance, a 3-dimensional architecture that stacks entangling and storage zones while preserving their size could be considered.
This may ensure reconfigurability on 2-dimensional intra-zone movements and simultaneously allows efficient inter-zone movement in a direction perpendicular to them.
However, potential thermal issues due to increased contact area between zones should be considered.

\section{Evaluation Methodology}

\begin{table}[h]
    \centering \footnotesize
    \renewcommand{\arraystretch}{1.00}
    \renewcommand{\tabcolsep}{1.6mm}
    \caption{Hardware Parameters for Neutral Atom Array}
    \label{t0}
    \begin{tabular}{|l||l|l|}
        \hline
        Pulse Time of 1Q Gate & \multicolumn{2}{l|}{625 ns (BB1 pulse, except RZ) \cite{bluvstein2022quantum}}  \\
        \hline
        Pulse Time of 2Q Gate & \multicolumn{2}{l|}{380 ns (global Rydberg pulse) \cite{bluvstein2022quantum}}  \\
        \hline
        Readout Imaging Time & \multicolumn{2}{l|}{500 $\mu$s \cite{bluvstein2024logical}}  \\
        \hline
        Fidelity of 1Q Gate & \multicolumn{2}{l|}{0.999 (Raman laser pulse \cite{evered2023high})}  \\
        \hline
        Fidelity of 2Q Gate & \multicolumn{2}{l|}{0.995 (time-optimal CZ \cite{evered2023high})} \\
        \hline
        Fidelity of Readout & \multicolumn{2}{l|}{0.998 (pushout and imaging \cite{evered2023high})}  \\
        \hline
        Trap Transfer Time & \multicolumn{2}{l|}{150 $\mu$s \cite{bluvstein2024logical}}  \\
        \hline
        Fidelity of Transfer & \multicolumn{2}{l|}{0.999 \cite{bluvstein2024logical}}  \\
        \hline
        Atom Array Size & \multicolumn{2}{l|}{41 x 41 (for each zone)}  \\
        \hline
        \# of Physical Qubits & \multicolumn{2}{l|}{1,681 qubits}  \\
        \hline
        \# of Logical Qubits & \multicolumn{2}{l|}{\textasciitilde 120 qubits (using EDFT \cite{bluvstein2024logical})} \\
        \hline
        AOD Trap Speed & \multicolumn{2}{l|}{0.55 $\mu$m/$\mu$s \cite{bluvstein2022quantum}}  \\
        \hline
        Shuttling (Unit Dist.) & \multicolumn{2}{l|}{21.8 $\mu$s}  \\
        \hline
        \multirow{2}{*}{Coherence Time} & \multicolumn{2}{l|}{100 s (in Storage zone) \cite{bluvstein2024logical},}  \\
        &  \multicolumn{2}{l|}{4 s (out of Storage zone) \cite{bluvstein2024logical}}  \\
        \hline
        Logical Basis & \multicolumn{2}{l|}{X-basis and Z-basis \cite{bluvstein2024logical}}  \\
        \hline
        Atom Unit Distance & \multicolumn{2}{l|}{12 $\mu$m (E. Zone), 6 $\mu$m (S. Zone) \cite{stade2024abstract}} \\
        \hline
        LD/ST Time & \multicolumn{2}{l|}{36.4 $\mu$s (Min.), 527.3 $\mu$s (Median)} \\
        \hline
        Fidelity Drop by X-talk & \multicolumn{2}{l|}{0.005 (1Q Gate) \cite{levine2019parallel}, 0.007 (CZ) \cite{levine2019parallel}} \\
        \hline
    \end{tabular}
\end{table}

\noindent
We build the in-house quantum program execution modeling for our evaluations, as there are currently no commercially available zoned neutral atom processor and no simulators for zoned architectures.
Hardware parameters for modeling the zoned neutral atom architecture are described in \cref{t0}.

\subsection{Modeling Zoned Rydberg Array Architecture}

\noindent
We assume a square-shaped array of atoms of 41 $\times$ 41 (1,681 physical qubits).
Although this array is larger than the size performed in the zoned architecture research \cite{bluvstein2024logical}, it is still less than the recently performed tweezer array of 2,088 physical qubits \cite{pichard2024rearrangement}.
The basic layout of the array of zoned architecture follows the original study: each zone is 20 micrometers apart from each other \cite{bluvstein2024logical}.
Each physical qubit is arranged in a two-dimensional lattice shape, 12 micrometers in the entangling zone and 6 micrometers in the storage zone.
The preparation of fault-tolerant qubits adopts the EDFT (post-selecting on flags and error detection) scheme \cite{bluvstein2024logical}, whereby 14 physical qubits constitute a single logical qubit.
Therefore, we could prepare 120 logical qubits under the assumption that 1,680 or more physical qubits are successfully trapped.

\subsection{Gate Operation Policies}

Three gate execution policies are considered for evaluations.

\textbf{Type 1} (\textit{default}): 
The execution of single-qubit and two-qubit gates is completely isolated according to the zone.
Qubits in the storage zone should be moved to the entangling zone when 2-qubit gate execution is required, and vice versa.

\textbf{Type 2}:
We consider a scenario in which each gate interleaving does not require inter-zone movement.
In this type, all qubits should start from the storage zone, but applying a local Raman laser (for the single-qubit gate execution) to the entangling zone is partially permitted.
Atoms subject to a single-qubit gate pulse should be sufficiently (more than 12 $\mu$m) apart from all the other atoms, which requires to the shuttling overhead within the entangling zone.
It is assumed that there is no fidelity drop due to the cross (X)-talk error when applying other kinds of pulses, thanks for the detuning techniques for protecting quantum states of qubits \cite{norcia2023midcircuit, lis2023midcircuit}.

\textbf{Type 3} (non-zoned architecture): 
It is assumed that both single-qubit and two-qubit gates can be executed in place on the single zone \cite{levine2019parallel}.
In this case, the atoms do not have to go out of the zone before the measurement.
Due to the in-place execution, we assume that atoms experience a fidelity drop by other kinds of pulse application, as shown in the \cref{t0}.

\subsection{Execution Time Estimation}

The following events are collected for run-time estimation.

\textbf{Load/Store}: 
It refers to the time for logical qubits to travel between entangling and storage zones.
Load and Store time includes for logical qubits to depart from their initial location and move to a specific location in another zone.
The AOD trap speed is limited to 0.55 $\mu$m/$\mu$s \cite{bluvstein2022quantum}. 
Therefore, the minimum Load or Store time is 36.4 $\mu$s, assuming that the distance between the entangling and storage zone is 20 $\mu$m.

\textbf{Trap Transfer}:
This means the time required for the handover of atoms between AOD traps and SLM traps.
Some logical qubits require a transfer between AOD and SLM traps due to the implementation of transversal entanglement gates or the movement constraints of AOD traps \cite{tan2024compiling, tan2024depth}.
The trap transfer time of approximately 150 to 300 $\mu$s is required \cite{bluvstein2024logical}.

\textbf{AOD Shuttling and SWAP}:
Note that AOD traps cannot cross each other and have several movement constraints \cite{bluvstein2024logical, tan2024compiling, stade2024abstract}.
The AOD trap movement policy for the baseline follows OLSQ-DPQA \cite{tan2024compiling}.
Assuming that the distance between each atom within the entangling zone is 12 $\mu$m and placed in a 2-dimensional square-shaped lattice structure, a qubit trapped in AOD requires about 21.8 $\mu$s to travel one edge of the unit lattice.
For zoned architectures, we modify the SWAP operation to always adopt a trap movement-based rather than a gate-based one.
This is because trap-movement-based SWAP is almost always more efficient than gate-based SWAP in zoned architectures, as discussed in the \cref{swap}.

\textbf{Readout} (Measurement): An imaging time of approximately 0.5 ms is required to measure the state of qubits \cite{bluvstein2024logical}.

\textbf{Error Correction}: 
This refers to the amount time it takes to perform the EDFT method \cite{bluvstein2024logical}, where it includes times for preparing logical qubits in $\lvert 0 \rangle$ or $\lvert + \rangle$ state through 7-qubit Steane code \cite{steane1996multiple}, applying parallel transversal entangling gates, and traveling ancilla logical qubits to the storage zone.

\textbf{Gate Exeution Time}:
The single-qubit pulse application time is assumed to be 625 nanoseconds \cite{wang2024atomique}.
This time is applied to single-qubit gates rotating arbitrary axes except for Z rotations.
The Z-rotation could be virtually implemented by the control software without consuming time \cite{bluvstein2024logical}.
The 2-qubit pulse time is assumed to be 380 nanoseconds \cite{wang2024atomique}.

\begin{table*}[h]
    \centering \footnotesize
    \renewcommand{\arraystretch}{1.00}
    \renewcommand{\tabcolsep}{1mm}
    \caption{Quantitative evaluations of Mantra over standard program execution on various scalable benchmarks.}
    \label{t1}
    \begin{tabular}{|c|c|c|c|c|c|c|c|c|c|c|c|c|}
        \hline
        \multirow{2}{*}{Workloads} & Logical & \multicolumn{3}{c|}{\# of LD/STs (X-basis calculation is allowed.)} & \multicolumn{3}{c|}{\# of Physical Gates} & \multicolumn{3}{c|}{Total Circuit Fidelity} \\ \cline{3-11}
         & Qubits  & Standard & Mantra & Reduced & Standard & Mantra & Reduced & Standard & Mantra & Improved \\ \hline \hline
        \multirow{3}{*}{GHZ} & 40Q & 78 & 10 (8) & 87.2\% (89.7\%) & 826 & 826 & 0.0\% & 0.75 & 0.76 & 1.5\% \\ 
        & 80Q & 158 & 2 (0) & 98.7\% (100.0\%) & 1,666 & 1,666 & 0.0\% & 0.56 & 0.57 & 2.6\% \\ 
        & 120Q & 238 & 2 (0) & 99.2\% (100.0\%) & 2,506 & 2,506 & 0.0\% & 0.42 & 0.43 & 3.1\% \\ \hline
        \multirow{3}{*}{PO} & 3Q & 36 & 6 & 83.3\% & 322 & 259 & 19.6\% & 0.89 & 0.89 & 0.9\% \\ 
        & 6Q & 180 & 6 & 96.7\% & 1,596 & 1,281 & 19.7\% & 0.55 & 0.58 & 4.6\% \\ 
        & 9Q & 432 & 6 & 98.6\% & 3,808 & 3,052 & 19.8\% & 0.24 & 0.27 & 11.4\% \\ \hline
        \multirow{3}{*}{QNN} & 8Q & 46 & 16 & 65.2\% & 959 & 567 & 40.9\% & 0.68 & 0.71 & 5.8\% \\ 
        & 16Q & 94 & 32 & 66.0\% & 3,731 & 2,051 & 45.0\% & 0.21 & 0.27 & 28.6\% \\ 
        & 24Q & 142 & 48 & 66.2\% & 8,295 & 4,431 & 46.6\% & 0.03 & 0.11 & 73.7\% \\ \hline
        \multirow{3}{*}{UCC} & 5Q & 160 & 40 & 75.0\% & 1,680 & 707 & 57.9\% & 0.57 & 0.65 & 14.9\% \\ 
        & 10Q & 360 & 40 & 88.9\% & 3,780 & 1,407 & 62.8\% & 0.28 & 0.40 & 40.0\% \\ 
        & 15Q & 560 & 40 & 92.9\% & 5,880 & 2,107 & 64.2\% & 0.14 & 0.24 & 71.5\% \\ \hline
        \multirow{3}{*}{VR} & 15Q & 84 & 84 & 0.0\% & 910 & 588 & 35.4\% & 0.74 & 0.78 & 4.7\% \\ 
        & 30Q & 174 & 174 & 0.0\% & 1,855 & 1,218 & 34.3\% & 0.54 & 0.59 & 9.7\% \\
        & 45Q & 264 & 264 & 0.0\% & 2,800 & 1,848 & 34.0\% & 0.39 & 0.45 & 14.6\% \\ \hline \hline
        \multicolumn{2}{|c|}{Geometric Mean} & 153.6 & 20.6 (18.5) & 86.6\% (87.9\%) & 2,010.1 & 1,297.8 & 35.4\% & 0.37 & 0.43 & 17.1\% \\ \hline
    \end{tabular}
\end{table*}

\subsection{High-Throughput Trap Movement Strategies}

Two strategies are applied to achieve high-throughput execution in zoned architectures.
They are implemented in our emulated modeling and are reflected in all runtime scenarios.

\textbf{Transferring Traps in Advance}:
The trap transfer time takes far longer than the gate pulse time.
We can hide the latency by performing trap transfers while logical qubits travel between zones.
Suppose one of the two logical qubits scheduled for a 2-qubit transversal operation is being loaded from storage zone, or it needs to go to the storage zone and come back. 
In that case, we transfer the remaining qubits waiting in the entangling zone to the SLM trap in advance.

\textbf{Placing Qubits near Zone Borders}:  
To minimize inter-zone travel time (LD/ST), physical qubits are arranged along the borders nearest to adjacent zones. 
Since the number of physical qubits is known before program execution, all qubits could be pre-placed as close to the zone borders as possible.

\subsection{Fidelity Estimation}

Each qubit's fidelity ($F$) can be estimated as follows:
$F = F_{1Q}^{N_{1Q}} \times F_{2Q}^{N_{2Q}} \times F_{\textrm{Transfer}}^{N_{\textrm{Transfer}}} \times F_{\textrm{Readout}} \times F_{\textrm{Decoherence}}$, 
where $F_{1Q}$ is the fidelity of a 1-qubit gate, $N_{1Q}$ is the number of applied 1-qubit gates, $F_{2Q}$ is the fidelity of a 2-qubit gate, $N_{2Q}$ is the number of applied 2-qubit gates, $F_{\textrm{Transfer}}$ is the fidelity of a trap transfer, $N_{\textrm{Transfer}}$ is the number of applied trap transfers, $F_{\textrm{Readout}}$ is the readout imaging fidelity, and $F_{\textrm{Decoherence}}$ is the fidelity by decoherence.
The gate fidelity itself is equivalent to each other in both non-zoned and zoned architectures since we consider homogeneous Rydberg atoms.
However, crosstalk errors by other types of gate pulses are considered as a fidelity drop term in the case of non-zoned architectures.
Decoherence follows a model that relaxes exponentially, where the time constant depends on whether each individual qubit exists in or out of storage zone: $F_{\textrm{Decoherence}} = e^{-(\tfrac{T_{\textrm{In}}}{100}+\tfrac{T_{\textrm{Out}}}{4})}$.
$T_{\textrm{In}}$ is the amount of time that the atom exists in the storage zone during the program execution, and $T_{\textrm{Out}}$ is the amount of time it exists out of the storage zone.
Total fidelity can be obtained by multiplying all fidelities from individual qubits.

\subsection{Quantum Program Benchmarks}

We utilize five scalable quantum circuits obtained from the MQT Bench \cite{quetschlich2023mqt}, SupermarQ \cite{tomesh2022supermarq} and QASMBench \cite{li2023qasmbench}: \texttt{GHZ} ( Greenberger–Horne–Zeilinger State Preparation), \texttt{PO} (Portfolio Optimization with QAOA), \texttt{QNN} (Quantum Neural Network), \texttt{UCC} (Unitary Coupled-Cluster Singles and Doubles ansatz randomly sampled), and \texttt{VR} (Vehicle Routing).
For $n$-qubit \texttt{UCC} circuits, it consists of 10 Pauli strings, where the $n$ matrices of each Pauli string are randomly sampled with the same probability among Pauli-I, X, Y, and Z.
Furthermore, we utilize six molecular simulation circuits for 12-qubit \texttt{LiH}, 14-qubit \texttt{BeH2}, 16-qubit \texttt{NH3}, 18-qubit \texttt{CH4}, 20-qubit \texttt{LiF}, and 22-qubit \texttt{MgH2}, where the Pauli strings of each molecule can be obtained from PySCF \cite{sun2018pyscf} and the circuit synthesis of the Jordan-Wigner transformation-based VQE circuits utilizes the Qiskit SDK \cite{qiskit2024}.
We also utilize Power Law \cite{hua2023caqr} and Sherrington-Kirkpatrick (SK) \cite{sherrington1975solvable} model graph-based QAOA for Max-Cut circuits with varying qubit and layer counts.
The gate cancellation is applied to all quantum benchmark programs by Qiskit Transpiler (optimization level = 3) \cite{qiskit2024}.

We consider execution per Pauli-string since we are interested in the scale of workloads that can be processed with fidelity that is meaningful to current zoned architectures, although estimating execution times for complete Pauli-strings is not challenging.
For example, the VQE circuit simulating \texttt{LiH} requires 36.0 minutes by NALAC and 11.6 minutes by \textit{Mantra}, which exceed the Rydberg atom's coherence time.


\section{Results and Analysis}

\subsection{Evaluations on Scalable Benchmarks} \label{scale}

\noindent
As shown in \cref{t1}, \textit{Mantra} reduces the number of LDs or STs, decreases physical gates, and improves circuit fidelity on average, compared to the standard execution.
\texttt{GHZ} 40Q differs in LD or ST counts from 80Q and 120Q since their CZ chain structures are different from each other, as discussed in \cref{efficiency}.
\texttt{GHZ}40Q has relatively low qubit requirements, allowing \textit{Mantra} to place logical qubits as thin rectangular-shaped arrays along the boundaries of other zones.
The parallel chain structure is more advantageous in this case.
For \texttt{GHZ} 80Q and 120Q, the fountain-shaped chain is more advantageous than the parallel ones due to further travel distance between zones.
Due to the proposed gate scheduling, all CZs in the \texttt{GHZ} 80Q and 120Q are processed in a single entangling zone execution step, requiring only two zone-to-zone movements. 
If X-basis initialization and measurement are permitted \cite{bluvstein2024logical}, H gates in GHZ circuits are unnecessary.
In this case, \texttt{GHZ} 80Q and 120Q can be performed without LD/ST. 

\texttt{PO} workloads require the number of LDs and STs of complexity proportional to the square of the number of qubits for standard execution (due to gate counts in the cost Hamiltonian), but \textit{Mantra} only requires 1 LD and 1 ST per QAOA layer, regardless of the number of qubits.
For workloads with deep circuit depth, such as \texttt{QNN}s and \texttt{UCC}s, the fidelity improvement by \textit{Mantra} is greater than for other workloads.
This is attributed to a reduction in the overall run-time by reducing LD and STs, resulting in less fidelity drop by the decoherence.
\textit{Mantra} does not always reduce LD/STs or gates.
For \texttt{GHZ} workloads where only the gate scheduling is applicable, the number of LD/STs is reduced, but that of the gates is not.
Since \texttt{VR} consists of fixed path-shaped CX chains with no synthetic flexibility, the number of LD/STs does not decrease.

\subsection{Evaluations on Molecular Simulation Workloads}

\begin{figure} [h] 
  \centerline {
  \includegraphics [width=\columnwidth] {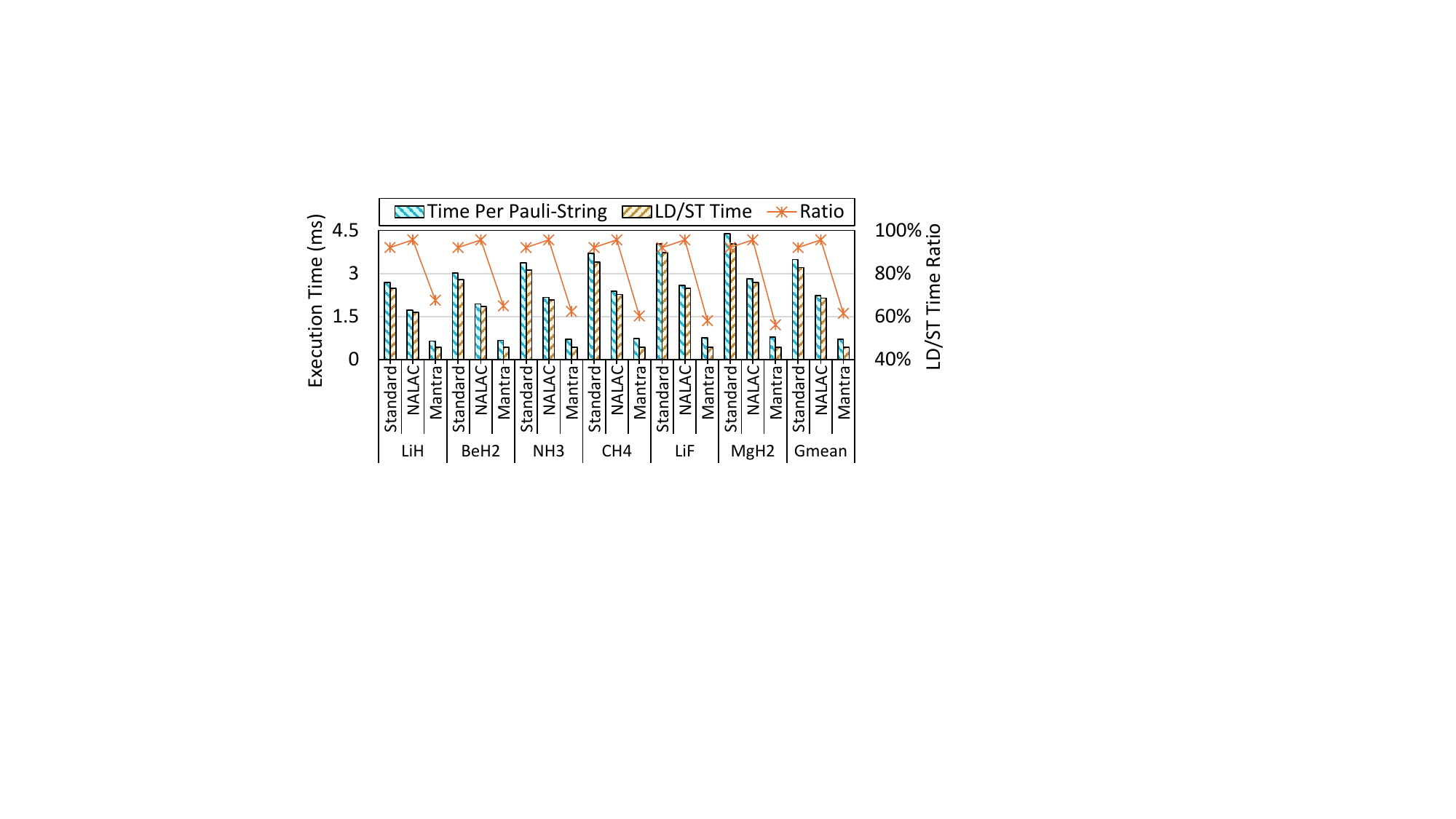} }
  \caption {
    Comparative analysis of average execution time per single Pauli-string of six VQE circuits using Jordan-Wigner encoder according to quantum program compilers.
  } 
  \Description[<short description>]{<long description>}
  \label{f11} 
\end{figure}

\noindent
\cref{f11} illustrates the total execution and LD or ST time per a single Pauli-string for VQE circuits according to the 3 execution scenarios: Standard approach, NALAC, and \textit{Mantra}.

As shown in \cref{f11}, both NALAC and \textit{Mantra} reduce execution time over the standard execution across all VQE circuits. 
NALAC reduces execution time per Pauli string by 36\% and the LD/ST time ratio by 33\% on average, over the standard execution. 
\textit{Mantra} reduces execution time per Pauli-string by 79\% and the LD/ST time ratio by 86\% on average, over the standard execution.
The performance improvement by \textit{Mantra} over NALAC can be attributed to its reduction of LD and ST operations, which are major bottlenecks in zoned architecture execution. 
This is also revealed by the ratio of LD/ST times to the entire program run time, which averages 96\% for NALAC, slightly higher than the standard execution ratio of 92\%, while \textit{Mantra} achieves a lower average of 61\%.
As the number of qubits for the molecular simulations increases, the execution reduction ratio of \textit{Mantra} against the standard program execution gradually increases because \textit{Mantra} requires a constant number of LD or ST instructions per single Pauli-string, regardless of the molecule workload.

\subsection{Evaluations on QAOA Workloads}

\begin{figure} [h] 
  \centerline {
  \includegraphics [width=\columnwidth] {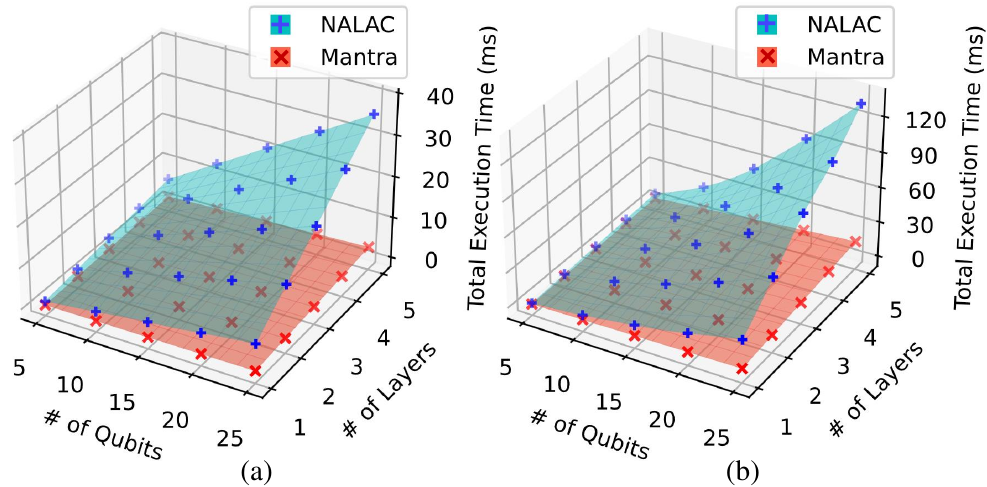} }
  \caption {
    A comparative analysis of the total execution time of QAOA (Quantum approximate optimization algorithm) circuits for the (a) Power Law models \cite{hua2023caqr} and (b) Sherrington-Kirkpatrick model \cite{sherrington1975solvable} as a target graph according to the number of qubits and layers: NALAC VS. \textit{Mantra}
    } 
  \Description[<short description>]{<long description>}
  \label{f12} 
\end{figure}

\noindent
\cref{f12} illustrates the total run-time of QAOA circuits for PL and SK models. 
In both compilers and target graphs, the execution time increases linearly with the number of QAOA layers. 
For PL graphs, the QAOA circuit depth scales linearly with the number of qubits. 
For SK models, the number of RZZs grows quadratically with the number of qubits. 
Although the depth complexity for SK models is linear due to the fermionic SWAP network \cite{hashim2022optimized}, the run-time increases nearly quadratically with the number of qubits due to the shuttling and trap transfer overheads in the entangling zone.

In NALAC, LDs or STs are required for the cost Hamiltonian, which scales with the number of RZZ operations.
However, \textit{Mantra} employs the new gate protocol, which eliminates the need to access the storage zone during processing of the cost Hamiltonian.
For PL models, the total execution time of \textit{Mantra} relative to NALAC decreases from 62\% to 89\% (average 79\%) in a given qubit range.
For SK models, the total execution time of \textit{Mantra} relative to NALAC decreases from 75\% to 91\% (average 86\%) in a given qubit range.
As the number of qubits and layers increases, \textit{Mantra} could further reduce the QAOA execution time over NALAC.
These results indicate that a major bottleneck in running QAOA circuits in zoned architectures may arise from inter-zoned movement overhead arising from processing the cost Hamiltonian, and \textit{Mantra} could efficiently reduce the inter-zoned movements.

\subsection{1Q Gate Execution Allowed in Entangling Zone}

One possible implementation of zoned architectures can permit 1-qubit gate execution within the entangling zones. 
Even in this case, atoms in the entangling zone would still need to be re-positioned to apply a local Raman beam realizing 1-qubit gates. 
In this scenario, while atoms requiring single-qubit gates no longer need to travel to the storage zone for each operation, the interleaving of single-qubit and two-qubit gate operations would still hinder program execution due to intra-zone atom movements. 
\textit{Mantra} can improve execution efficiency by effectively mitigating trap transfer and shuttling overheads, even under this operation policy.

\begin{figure} [h] 
  \centerline {
  \includegraphics [width=\columnwidth] {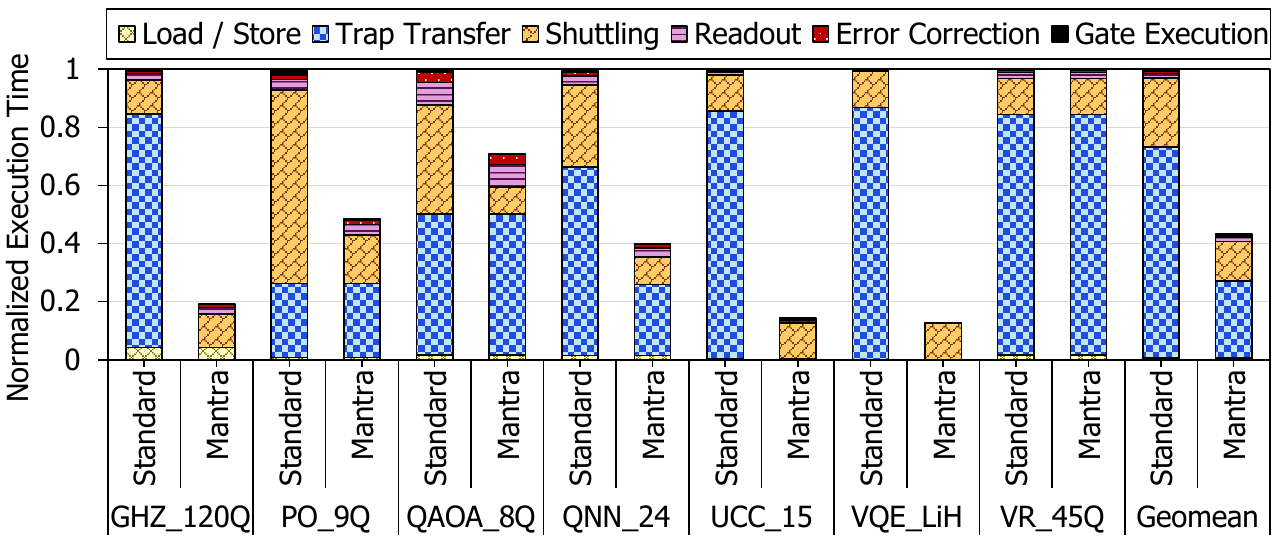} }
  \caption {
    Normalized execution time comparison for various workloads by Standard and \textit{Mantra} when 1Q gates are allowed in the entangling zone (i.e., Type 2 operation policy).
  } 
  \Description[<short description>]{<long description>}
  \label{f125} 
\end{figure}

This section evaluates the performance achieved by \textit{Mantra} under the Type 2 policy, where the Raman laser is partially allowed in the entangling zone. 
\cref{f125} illustrates the normalized run-times for various workloads under this policy, comparing the Standard and \textit{Mantra} compilation strategies. 
Allowing 1-qubit gates in the entangling zone suppresses the \texttt{Load/Store} overhead remarkably, as atoms no longer need to move between the storage and entangling zones for each gate interleaving. 
While \texttt{Load/Store} are the primary bottlenecks under the Type 1 policy, they account for only 0.7\% of the total run-time in this case. 
Instead, \texttt{Trap Transfer} and \texttt{Shuttling} emerge as dominant execution time bottlenecks.

The program rewriting strategies by \textit{Mantra} can reduce Trap Transfer and Shuttling overheads.  
For instance, in workloads such as \texttt{QAOA\_8Q}, \texttt{UCC\_15}, and \texttt{VQE\_LiH}, \textit{Mantra} eliminates the need for Trap Transfer, unlike the Standard approach.  
Across seven workloads, the geometric mean of Trap Transfer execution overhead is reduced by approximately 64\%.
Similarly, in workloads such as \texttt{PO\_9Q}, \texttt{QAOA\_8Q}, and \texttt{QNN\_24}, \textit{Mantra} efficiently reduces Shuttling time compared to the Standard.  
On the geometric mean across all workloads, Shuttling is reduced by approximately 43\%.  
When these improvements are combined, the overall execution is reduced by approximately 57\% on the geometric mean across all workloads.
These results demonstrate that \textit{Mantra} not only reduces zone-to-zone movement but also effectively suppresses execution overhead in scenarios where one-qubit and two-qubit gate operations occur within the same zone.

\subsection{Evaluations on Zoned/Non-Zoned Architectures}

\noindent
We compare the performance of different compilers on zoned (Type 1) and non-zoned (Type 3) gate operational policies.

\begin{figure} [h] 
  \centerline {
  \includegraphics [width=\columnwidth] {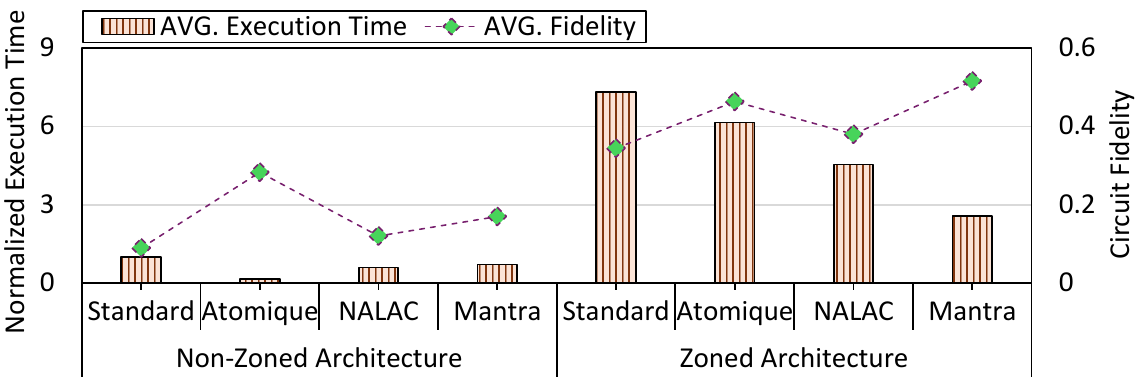} }
  \caption {
    Comparison of normalized average execution time and fidelity of scalable benchmarks under different compilers in non-zoned and zoned neutral atom architectures (lower execution time is better and higher fidelity is better).
  } 
  \Description[<short description>]{<long description>}
  \label{f13} 
\end{figure}

\textbf{On the architectural adoption perspective}:
Zoned architectures can achieve better fidelity than non-zoned architectures by isolating gate execution zones. 
Unfortunately, they may require longer execution times due to the overhead associated with inter-zone movement. 
As shown in \cref{f13}, results indicate that zoned architectures can take an average of 7.3$\times$ longer in standard execution compared to non-zoned architectures.
This increased execution time may raise concerns for those considering the adoption of zoned architectures. 
Nevertheless, we argue that achieving high-accuracy computational results is often more critical, even if it comes at the cost of longer execution times.
\textit{Mantra} supports this argument by reducing the execution time of zoned architectures by 65\% compared to standard execution. 
Although the execution time with \textit{Mantra} on zoned architectures still remains longer than that of non-zoned architectures, it achieves a reasonable runtime of 2.6$\times$ and far improves program fidelity, increasing it from 9\% to 52\% on average over standard execution in non-zoned architectures.

\textbf{On the domain-specific compiler perspective}:
Unlike non-zoned architectures, zoned architectures pose a new execution bottleneck (i.e., inter-zone movement), which may require a different feature for the compiler to mitigate them. 
\textit{Mantra} focuses on reducing this inter-zone travel overhead, which can further reduce the overall run-time in the zoned architecture compared to state-of-the-art compilers for neutral atom-based quantum computers.
Atomique \cite{wang2024atomique}, which is a general-purpose atom array compiler, provides significant performance benefits in various neutral atom array architectures.
In non-zoned architectures, Atomique \cite{wang2024atomique} achieves the shortest average execution time, about  7$\times$ faster than the standard execution, and provides the highest circuit fidelity among compilers.
Atomique \cite{wang2024atomique} leverages the high-parallelism mapper and router to eliminate shuttling and trap transfer significantly, but this overhead may not be a bottleneck in zoned architectures as described in \cref{f3}.
As described in \cref{f13}, it is observed that the run-time reduction ratio by Atomique \cite{wang2024atomique} is less for zoned architectures than for non-zoned architectures.
NALAC \cite{stade2024abstract} is the first and latest compiler for zoned architectures, and it efficiently performs gate execution by leveraging DSatur algorithm-based parallel AOD shuttling.
The underlying idea of NALAC \cite{stade2024abstract} mainly considers gate scheduling within entangling zones, not reducing the number of inter-zone travels (note that the LD/ST time itself is reduced by 40\% due to the AOD shuttling parallelism.). 
Thus, NALAC may provide fewer execution time reductions in the zoned architecture than the \textit{Mantra}.

\subsection{Execution Breakdown According to Compiler}

\begin{figure} [h] 
  \centerline {
  \includegraphics [width=\columnwidth] {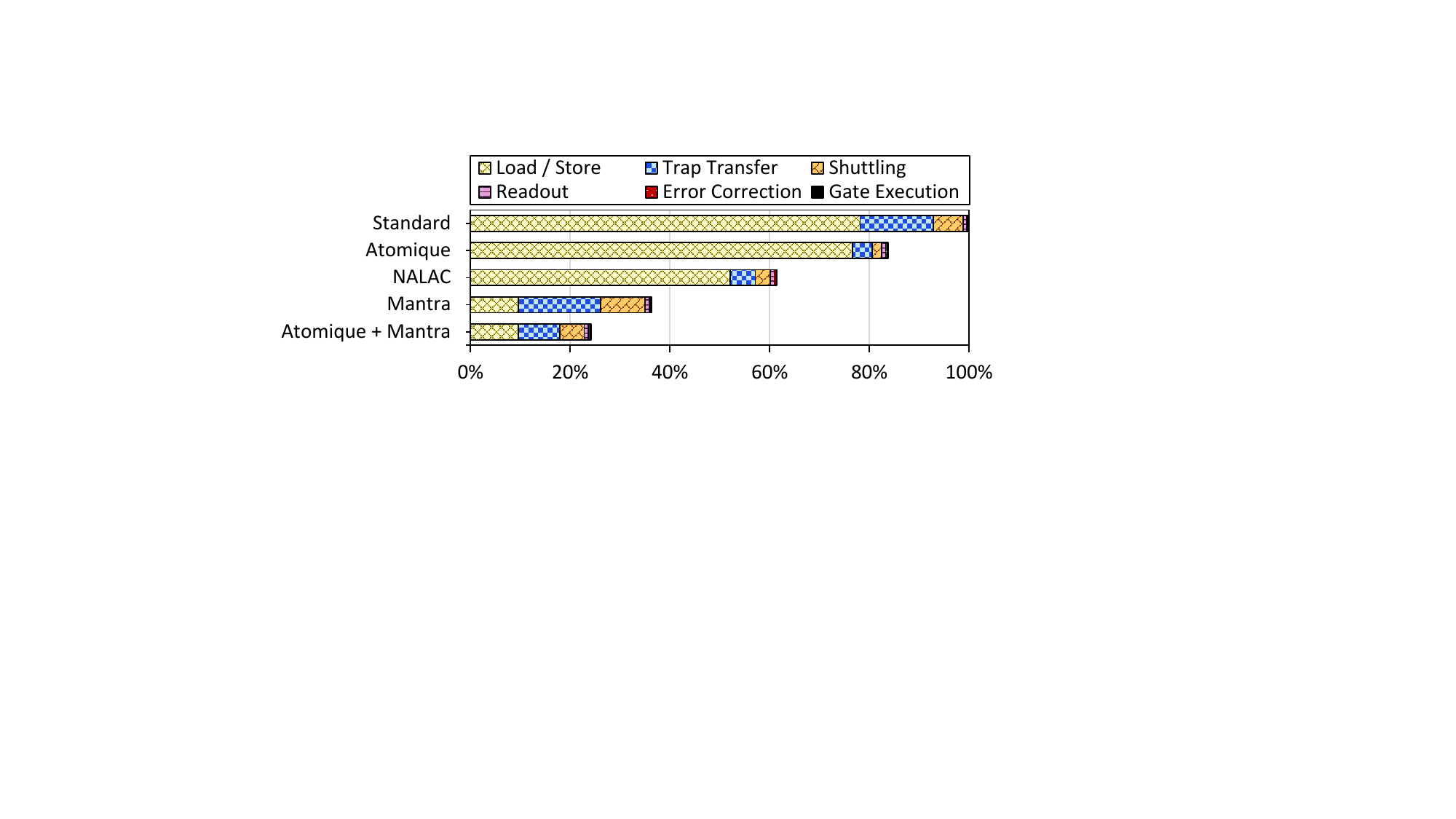} }
  \caption {
    Execution time breakdown analysis with different quantum compilers in the zoned neutral atom architecture.
  } 
  \Description[<short description>]{<long description>}
  \label{f14} 
\end{figure}

\noindent
This section analyzes the execution time breakdown according to different compilers on zoned architectures. 
As shown in \cref{f14}, \textit{Mantra} reduces the average execution time by 64\% compared to standard execution, 57\% compared to Atomique, and 41\% compared to NALAC. 
\textit{Mantra} achieves the lowest average execution time by reducing LD/ST more than previous compilers, but it also increases trap transfer and shuttling time. 
This increase can be attributed to revealing the overhead of trap transfer and shuttling time, which was previously hidden within LD/ST operations, as \textit{Mantra} reduces inter-zone movements.
In standard execution, while some logical qubits are moved to the storage zone for single-qubit gate execution, other qubits remaining in the entangling zone can perform AOD to SLM transfers in advance. 
These preparations for later transversal entangling operations are hidden within the inter-zone travel time.
By eliminating the need for certain qubits to move to the storage zone, \textit{Mantra} unveils these shuttling and trap transfer times. 
While \textit{Mantra} primarily focuses on reducing (inter-zone) LD/ST overhead, Atomique targets (intra-zone) trap transfer and shuttling optimization. 
Combining both methods could further enhance the performance of zoned architectures, as shown in \cref{f14}.

\section{Related Work}

This section discusses the latest related studies.
The topics below relate to the techniques or schemes utilized by \textit{Mantra}.

Recent studies discuss compilation methodologies to efficiently upload Hamiltonian simulation circuits to super-conductor-based quantum processors.
\emph{Paulihedral} \cite{li2022paulihedral} introduces a compiler framework that optimizes quantum simulation kernels by utilizing a Pauli intermediate representation (Pauli-IR).
\emph{Paulihedral} performs a compilation process in Pauli string-level representations instead of gate-based circuit level, resulting in lower transformation overhead.
\emph{Tetris} \cite{jin2024tetris} proposes a framework for VQA using an intermediate representation (Tetris-IR) based on Pauli strings.
\emph{Tetris} efficiently cancels CXs and minimizes SWAP insertion overhead.

Alternative gate templates \cite{jang2023quixote} could be applied depending on qubit architecture or input quantum states to reduce program execution costs.
\emph{Orchestrated Trios} \cite{duckering2021orchestrated} introduces an alternative strategy for the Toffoli gate decomposition.
\emph{Orchestrated Trios} \cite{duckering2021orchestrated} adaptively leverages two templates for the Toffoli gate: one that uses 6 CXs, requiring connections between all pairs of 3 qubits, and an alternative that uses 8 CXs, connecting only two pairs among the three qubits.
\emph{Relaxed Peephole Optimization} (RPO) \cite{liu2021relaxed} replaces some gates with an alternative one when its inputs are in certain states.

The order of gate execution could be rearranged to improve the accuracy or speed of quantum hardware. 
To reduce program depth, the \emph{Intelligent Approach} \cite{alam2020circuit} modifies the order of ZZ-rotation operations that configure the QAOA cost Hamiltonian. 
\emph{VAQEM} \cite{ravi2022vaqem} adjusts the gate alignment to execute some gates at a timing that minimizes the impact of idling errors.
While each approach aims to reduce circuit depth and minimize error from idling, the gate alignment of \textit{Mantra} focuses on reducing inter-zone movements of qubits.

\section{Conclusion}
Inter-zone travels of qubits may be an execution bottleneck on zoned architecture.
This may be intensified by existing program structures configured to frequently interleaved gate executions.
Na\"ively applying the existing program structure to the zoned architecture could require quite inter-zone travels.
Using several case studies, we explain program rewriting strategies (\textit{Mantra}) to reduce inter-zone atoms' movements.

\begin{acks}
The authors extend appreciation to the reviewers for the feedback. 
The authors are grateful to Hengyun Zhou for discussions on the real zoned neutral atom processor, Joonhee Choi for discussions on the tweezer array, and Young Jung Kang for discussions on the quantum error correction. 
This research was funded by the National Research Foundation of Korea (NRF), supported by the Korean government under the project ``Creation of the Quantum Information Science R\&D Ecosystem Based on Human Resources'' (No. RS-2023-00303229). 
Won Woo Ro is the corresponding author and the correspondences of this research should be directed to him.
\end{acks}






\appendix
\section{Preparations for Fault-Tolerant Qubits} \label{appendixa}

\noindent
The fault-tolerant quantum computing aims to protect quantum information in qubits from errors caused by noise and decoherence \cite{peres1985reversible}.
Quantum error correction (QEC) is required to achieve such fault-tolerant quantum computing, and various error correction codes such as Steane code \cite{steane1996multiple, goudarzi2014design}, surface code \cite{bluvstein2022quantum, gottesman1997stabilizer, viszlai2023architecture}, toric code \cite{kitaev1997quantum}, and 3D block code \cite{bluvstein2024logical} could be applied to neutral atom-based quantum computers.

\begin{figure} [h] 
  \centerline {
  \includegraphics [width=\columnwidth] {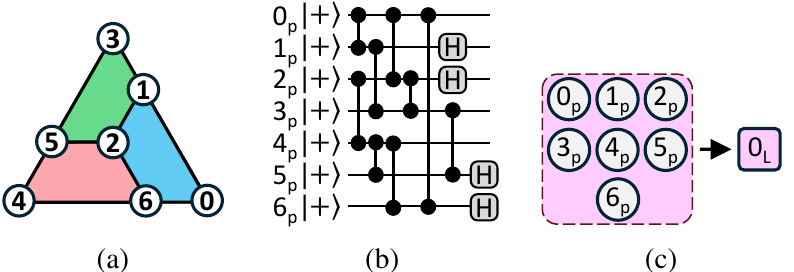} }
  \caption {
    (a) A graph-based representation of 7-qubit Steane code for error correction, 
    (b) A quantum circuit to initialize fault-tolerant logical qubits using Steane code, and
    (c) A single logical qubit ($0_L$) consisting of 7 physical qubits.
  } 
  \Description[<short description>]{<long description>}
  \label{f0} 
\end{figure}

Among these QEC codes, we adopt the 7-qubit Steane code \cite{steane1996multiple} as shown in \cref{f0}, which is relatively straightforward to prepare fault-tolerant qubits and allows to achieve transversal operations in full Clifford gate groups \cite{bluvstein2022quantum}.
\cref{f0} (a) represents a graph-based representation of the 7-qubit Steane code, where the colors in the graph represent stabilizer plaquettes that detect errors by the measurement results from seven physical qubits \cite{nigg2014quantum, ryan2021realization}.
\cref{f0} (b) shows a quantum circuit to prepare a logical state of $\lvert + \rangle (= (\lvert 0 \rangle + \lvert 1 \rangle)/\sqrt{2})$ using the Steane code \cite{bluvstein2022quantum}.
This work describes seven physical qubits (circular shaped) encoded by the Steane code as a single logical qubit (square shaped), as shown in \cref{f0} (c).

\section{Hamiltonian for Time-Optimal Gates} \label{appendixb}

The implementation of entangling gates, such as Controlled-Phase (\texttt{CPHASE} \cite{evered2023high}) and Levine-Pichler (LP \cite{jandura2022time}) gates, leverages the Rydberg blockade effect and van der Waals interactions \cite{bluvstein2022quantum} in neutral atom quantum processors. 
The Hamiltonian governing these gates could be expressed as follows.
\begin{align*}
  H(t) &= \sum_{i} \frac{\Omega_i(t)}{2} \big( \sigma^+_i + \sigma^-_i \big) + \sum_{i} \Delta_i n_i + \sum_{i<j} V_{ij} n_i n_j,
\end{align*}

The parameters in the Hamiltonian are defined as follows \cite{walker2012entanglement}. 
$\Omega_i(t)$ represents the Rabi frequency controlling the transition between the ground state $\lvert g \rangle$ and the Rydberg-excited state $\lvert r \rangle$. 
The detuning $\Delta_i$ is the laser offset from resonance. 
The operator $n_i = \lvert r \rangle \langle r \rvert$ is the Rydberg population operator, indicating whether the atom is in the Rydberg state. 
The interaction strength between two atoms $i$ and $j$ is given by $V_{ij} = C_6 / r_{ij}^6$, which describes van der Waals interaction as a function of the distance $r_{ij}$ \cite{heilbron1969genesis}. 
$\sigma^+_i$ and $\sigma^-_i$ are the raising and lowering operators for the transitions of atom $i$.

The van der Waals interaction term \( V_{ij} n_i n_j \) is critical for inducing phase shifts necessary for entangling operations. 
By modulating the Rabi frequency \( \Omega(t) \) and detuning \( \Delta(t) \), the dynamics of the system can be controlled to implement gates with high fidelity and minimal duration \cite{norcia2023midcircuit, lis2023midcircuit}. 
For the \texttt{CPHASE} gate, the Rabi frequency and detuning govern the phase accumulation on the $\lvert 11 \rangle$ state. 
Similarly, LP gate applies controlled phase shifts, assigning the same specific phase to \(\lvert 01 \rangle\) and \(\lvert 10 \rangle\) while applying a different one to \(\lvert 11 \rangle\).

\section{Proposed Arbitrary ZZ-Rotation Protocol} \label{appendixc}

This section discusses a unitary that combines time-optimal \texttt{CPHASE} with LP gate to realize  ZZ-interaction operations with arbitrary angles.
The target unitary for the $RZZ(\gamma)$ is:
\begin{align*}
  RZZ(\gamma) &=
  \begin{bsmallmatrix}
        1 & 0 & 0 & 0 \\
        0 & e^{i\gamma} & 0 & 0 \\
        0 & 0 & e^{i\gamma} & 0 \\
        0 & 0 & 0 & 1 \\
  \end{bsmallmatrix}.
\end{align*}

It is realized by combining 2 gates. 
A \texttt{CPHASE} is shown as:
\begin{align*}
  U_{\text{CPHASE}}(\phi) &=
  \begin{bsmallmatrix}
        1 & 0 & 0 & 0 \\
        0 & 1 & 0 & 0 \\
        0 & 0 & 1 & 0 \\
        0 & 0 & 0 & e^{i\phi} \\
  \end{bsmallmatrix},
\end{align*}
which applies a phase $\phi$ to the $\lvert 11 \rangle$.
The \texttt{CPHASE}($\phi$) gate is implemented using a time-optimal protocol, where the Rabi frequency \(\Omega(t)\) and detuning \(\Delta(t)\) are modulated to achieve the desired phase \(\phi\) on the \(\lvert 11 \rangle\) state \cite{evered2023high}. 
The phase accumulation is given by \(\phi(t) = \phi_0 + \tfrac{\Omega^2}{\Delta} t\), where \(\phi_0\) is the initial phase offset. 
To shape the pulse, the Rabi frequency is adjusted as \(\Omega(t) = \Omega_{\text{max}} \exp\big(-\tfrac{t^2}{2\sigma^2}\big)\), where \(\sigma\) determines the width of the pulse, and \(\Omega_{\text{max}}\) is the maximum amplitude \cite{evered2023high}. 
This smooth Gaussian profile minimizes off-resonant scattering and enhances gate fidelity.
The gate duration is determined by \(\tau_{\text{CPHASE}} = \tfrac{2\pi}{\sqrt{\Delta^2 + \Omega_{\text{max}}^2}}\), where \(\Delta\) is the detuning, and \(\Omega_{\text{max}}\) is optimized to ensure sufficient phase accumulation while minimizing \(\tau_{\text{CPHASE}}\) \cite{evered2023high}.
An LP gate can be represented as:
\begin{align*}
  U_{\text{LP}}(\gamma) &=
  \begin{bsmallmatrix}
        1 & 0 & 0 & 0 \\
        0 & e^{i\gamma} & 0 & 0 \\
        0 & 0 & e^{i\gamma} & 0 \\
        0 & 0 & 0 & e^{i(2\gamma + \pi)} \\
  \end{bsmallmatrix},
\end{align*}
which applies a phase $\gamma$ to the $\lvert 01 \rangle$ and $\lvert 10 \rangle$ states, and $2\gamma + \pi$ to the $\lvert 11 \rangle$.
When combined, the resulting unitary becomes:
\begin{align*}
  U &= U_{\text{LP}}(\gamma) \cdot U_{\text{CPHASE}}(\phi) =
  \begin{bsmallmatrix}
        1 & 0 & 0 & 0 \\
        0 & e^{i\gamma} & 0 & 0 \\
        0 & 0 & e^{i\gamma} & 0 \\
        0 & 0 & 0 & e^{i(2\gamma + \pi + \phi)} \\
  \end{bsmallmatrix}.
\end{align*}

To cancel the additional phase term introduced by the LP gate and achieve the desired \(RZZ(\gamma)\) operation, the \texttt{CPHASE} gate is configured to apply \(\phi_{\text{CPHASE}} = -2\gamma - \pi\), thereby ensuring the total phase shift on \(\lvert 11 \rangle\) satisfies \(2\gamma + \pi + \phi = 0\). 

This proposed gate protocol is expected to improve program execution efficiency not only in the zoned neutral atom architecture but also in the non-zoned architecture.
By adopting the time-optimal LP gate \cite{jandura2022time} and the \texttt{CPHASE}($\phi$) gate \cite{bluvstein2024logical, evered2023high}, \textit{Mantra} can realize arbitrary rotating ZZ-interaction gates with shorter pulse duration than existing CX-based decomposition in both types of neutral atom architectures.

\balance
\bibliographystyle{ACM-Reference-Format}
\bibliography{sample-base}

\end{document}